\documentclass[
reprint,
superscriptaddress,
footinbib,
bibnotes,
amsmath,amssymb,
longbibliography,
aps,
pra,
]{revtex4-2}
\usepackage[ngerman,english]{babel}
\usepackage{graphicx}
\usepackage{dcolumn}
\usepackage{braket}
\usepackage{textcomp, gensymb}
\usepackage{bm}
\usepackage{amsmath}
\usepackage[dvipsnames]{xcolor}
\usepackage[utf8]{inputenc}
\usepackage[T1]{fontenc}
\usepackage{csquotes}
\usepackage{thmtools}
\usepackage{url}

\usepackage{breakurl}
\usepackage[normalem]{ulem}
\usepackage[breaklinks]{hyperref}
\hypersetup{pdfborder=0 0 0}

\usepackage[version=4]{mhchem}
\usepackage{dsfont}
\usepackage{natbib}
\usepackage{empheq}
\usepackage{cases}
\usepackage[all]{hypcap}
\usepackage{siunitx}
\usepackage[capitalize]{cleveref}
\usepackage{placeins}
\usepackage{listings}
\usepackage{tabularx}
\usepackage{nicematrix}
\usepackage{stmaryrd}
\usepackage{footmisc}
\usepackage{tikz}
\usetikzlibrary{quantikz2}
\usepackage{printlen}
\usepackage{ulem}
\usepackage{xfp}
\usepackage{array}
\usepackage{booktabs}

\usetikzlibrary{shapes, arrows}

\newcommand{\TimeSQEbenchmark}{446--3212}
\newcommand{\TimeDQEbenchmarkLow}{2549}
\newcommand{\TimeDQEbenchmarkHigh}{14189}
\newcommand{\TimeDQEbenchmark}{\TimeDQEbenchmarkLow--\TimeDQEbenchmarkHigh}
\newcommand{\improvementFactorSQE}{1.5--11.1}
\newcommand{\improvementFactorDQE}{2.7--15.3}
\newcommand{\improvementUpTo}{$15.3$}

\newcommand{\TimeSQEpulsebased}{289}
\newcommand{\TimeDQEpulsebased}{927}

\newcommand{\pih}{\frac{\pi}{2}}

\newcommand{\Xpih}{R_x (\pih)}
\newcommand{\VZ}{R_z (\theta)}

\begin{document}

\title{Pulse-optimised circuit elements for scalable and noise-resilient quantum chemistry}
\author{Henrik Gothen}
\affiliation{Hitachi  Cambridge  Laboratory,  J.  J.  Thomson  Ave., Cambridge,  CB3  0US,  United  Kingdom
}
\affiliation{ Cavendish Laboratory,  Department of Physics,  University  of  
Cambridge,  Cambridge,  CB3  0US,  United  Kingdom
}
\author{Christopher K. Long}
\affiliation{Hitachi  Cambridge  Laboratory,  J.  J.  Thomson  Ave., Cambridge,  CB3  0US,  United  Kingdom
}
\affiliation{ Cavendish Laboratory,  Department of Physics,  University  of  
Cambridge,  Cambridge,  CB3  0US,  United  Kingdom
}
\author{Djamila Hiller}
\affiliation{ Cavendish Laboratory,  Department of Physics,  University  of  
Cambridge,  Cambridge,  CB3  0US,  United  Kingdom
}
\author{Yunming Qian}
\affiliation{ Cavendish Laboratory,  Department of Physics,  University  of  
Cambridge,  Cambridge,  CB3  0US,  United  Kingdom
}
\author{Crispin H. W. Barnes}
\affiliation{ Cavendish Laboratory,  Department of Physics,  University  of  
Cambridge,  Cambridge,  CB3  0US,  United  Kingdom
}
\author{Normann Mertig}
\affiliation{Hitachi  Cambridge  Laboratory,  J.  J.  Thomson  Ave., Cambridge,  CB3  0US,  United  Kingdom
}
\author{David R. M. Arvidsson-Shukur}
\affiliation{Hitachi  Cambridge  Laboratory,  J.  J.  Thomson  Ave., Cambridge,  CB3  0US,  United  Kingdom
}
\date{\today}

\begin{abstract}
Useful chemistry calculations on near-term quantum processors are hindered by current algorithmic runtimes. 
We develop a methodology to significantly reduce these runtimes. Typically, variational-quantum-eigensolver (VQE) algorithms are implemented as sequences of primitive gates.
Our methodology, instead, relies on gradient-ascent pulse-engineered optimisation to construct hardware-tailored pulses for the direct implementation of VQEs.
As problem sizes increase, it quickly becomes intractable to optimise a pulse that implements an entire VQE ansatz circuit. However, leading VQEs are constructed in a modular fashion. A problem-tailored VQE is assembled from parameterised circuit elements that simulate hopping between two or four electronic spin orbitals. We show that these circuit elements can be implemented more efficiently using hardware-tailored pulses.
We numerically showcase our methodology on a silicon spin-qubit quantum processor. We find that common circuit elements, known as single- and double-qubit excitations, can be implemented in less than \TimeSQEpulsebased~ns and \TimeDQEpulsebased~ns, respectively. Compared with conventional gate-based implementations, our pulse-accelerated qubit excitations provide a scalable approach for faster---and thus more noise-robust---quantum-chemistry simulations by cutting the VQE runtimes by up to a factor of \improvementUpTo. 
\end{abstract}
\maketitle


\section{Introduction}
Calculations of molecular energy levels constitute a crucial but difficult component of computational chemistry. Current classical-computing methods are restricted either to small molecules, where exact solutions are feasible~\cite{ROSSI1999530}, or to large molecules, where approximate algorithms (\textit{e.g.,} density functional theory) perform well~\cite{Becke2014}. The important class of strongly correlated molecules with system sizes of around $50$--$200$ spin orbitals lies outside the regime of classical simulability. Quantum computers may offer a potential route to efficiently simulate this class of molecules~\cite{Feynman1982Jun, Guzik05, Cao2019, McArdle20}. However, insufficient coherence times and scalability issues have precluded implementations on present quantum-computing hardware~\cite{Dalton2024, TILLY20221}. 

The variational quantum eigensolver (VQE) is a hybrid quantum--classical algorithm, tailored to perform molecular simulations on imperfect hardware~\cite{peruzzo2014originalVQE, GrimsleyAdapt}. 
In its basic form, the VQE estimates the ground-state energy $E_0$ of a molecular Hamiltonian $H$.
To warm start this estimation, an input wavefunction $\ket{\Psi_0}$ is prepared in accordance with some classical ground-state approximation (often the Hartree--Fock state). A shallow parameterised quantum circuit prepares and measures the output state $| \Psi(\vec{\theta}) \rangle = U(\vec{\theta}) \ket{\Psi_0}$ and thus enables the calculation of an energy-expectation value $E(\vec{\theta}) = \langle \Psi(\vec{\theta}) | H | \Psi(\vec{\theta}) \rangle$.  
Guided by a classical parameter optimiser, the VQE iteratively minimises the energy expectation. Thus, the energy expectation tends to the ground-state energy via the Rayleigh--Ritz variational principle: $E(\vec{\theta}) \geq E_0$.

The last decades have seen the development of a plethora of VQE techniques~\cite{peruzzo2014originalVQE,Ramoa2025,Yordanov_2021,Anastasiou2024,Long24,TILLY20221,Tang2021,Zhang2021,Gomes2021,Yordanov2022eQEBAdapt,OverlapAdaptVQE_2023,Shkolnikov2023avoidingsymmetry,Fitzpatrick2024,Raffaele2018,Ryabinkin2018,Romero_2019,Lee2019,zhang2020variationalquantumeigensolversvariance,Ryabinkin2020iQCCVQE,Zhang2022ClusterVQE,Burton2023DISCOVqe,FoldedSpectrumExcStateVQE_2024,GrimsleyAdapt,Ibrahim2022,Araz2025,Meitei2021,Asthana2023,Egger2023,Meirom2023,Sherbert2025,Long2024MinEvolTimes}. Modern VQEs adaptively build molecule-tailored circuits from predefined \textit{circuit elements}~\cite{GrimsleyAdapt,Tang2021,Yordanov_2021,Zhang2021,Gomes2021,Yordanov2022eQEBAdapt,OverlapAdaptVQE_2023,Shkolnikov2023avoidingsymmetry,Fitzpatrick2024,Anastasiou2024,Long24,Ramoa2025,Yordanov_2020}. The most promising circuit elements satisfy two criteria: physical relevance and hardware efficiency. Such circuit elements simulate the wavefunction transfer between two or four electron-spin orbitals using few primitive single- and two-qubit gates~\cite{Yordanov_2020}. Currently, useful VQE simulations of unknown molecules lie outside the reach of state-of-the-art hardware. The bottleneck is decoherence: despite the successful efforts in reducing gate counts in VQEs, their circuits are too long to withstand the noise of present-day devices~\cite{Dalton2024}.

\begin{table}[tb]
\renewcommand{\arraystretch}{1.4}
\centering
\begin{ruledtabular}
\begin{tabular}{cccc}
\textbf{Time} &
\textbf{Gate-based} &
\textbf{Pulse-based} &
\textbf{Acceleration} \\
\colrule
$T_{\mathrm{SQ}}$ &
$\geq \TimeSQEbenchmark\,\mathrm{ns}$ &
$\leq \TimeSQEpulsebased\,\mathrm{ns}$ &
$\geq \improvementFactorSQE$ \\
$T_{\mathrm{DQ}}$ &
$\geq \TimeDQEbenchmark\,\mathrm{ns}$ &
$\leq \TimeDQEpulsebased\,\mathrm{ns}$ &
$\geq \improvementFactorDQE$
\end{tabular}
\end{ruledtabular}
\caption{\textbf{Execution times $T_{\mathrm{SQ}}$ and $T_{\mathrm{DQ}}$ of single- and double-qubit excitations}
implemented via gate- and pulse-based approaches, respectively, on a silicon device.
The execution times of the gate-based approach depend on the detuning; hence execution times are reported as ranges.
The last column summarises the acceleration factors.
}\label{tab:mainresultsconcise}
\end{table}

One way to mitigate this problem is to replace gate circuits with pulse-based (Hamiltonian) methods~\cite{Li2017,Lu2017,Yang2017,lloyd2020quantumpolardecompositionalgorithm,Magann2021,lloyd2021hamiltoniansingularvaluetransformation,Meitei2021,Ebadi2022,Ibrahim2022,Asthana2023,Egger2023,Meirom2023,Araz2025,Sherbert2025,Long2024MinEvolTimes}. In gate-based quantum computing, an algorithm is transpiled into a circuit of primitive gates. The hardware's control pulses are then optimised and compiled to implement each of these gates. In pulse-based quantum computing, instead, the hardware pulses are optimised to implement the whole or parts of the algorithm directly. Thus, pulse-based quantum computing enables fine-grained hardware control and may cut the temporal overhead of compilation and transpilation.  Previous works have explored the prospect of pulse-based VQE algorithms on superconductor~\cite{Meitei2021,Ibrahim2022,Asthana2023,Egger2023,Meirom2023,Araz2025,Sherbert2025}, semiconductor~\cite{Long2024MinEvolTimes}, nuclear-magnetic-resonance~\cite{Li2017,Lu2017}, and neutral-atom~\cite{Ebadi2022} hardware.
Proof-of-principle simulations have demonstrated that pulse-based VQEs can be considerably faster than gate-based ones in the calculation of small-molecule energies~\cite{Meitei2021,Asthana2023,Egger2023,Meirom2023,Sherbert2025,Long2024MinEvolTimes}. 
Thus, pulse-based methods may bring quantum simulation of useful molecules within the coherence times of near-term hardware. However, while leading gate-based VQE algorithms have shown strong indications of being scalable to larger systems, current pulse-based VQE algorithms have not. The pulse optimisation quickly becomes intractable with increasing problem size~\cite{Sherbert2025}; there is little prospect of designing pulses for molecules with tens of spin orbitals. A natural question is: Can one combine the scalability of leading gate-based VQEs~\cite{GrimsleyAdapt,Yordanov_2021,Ramoa2025} with the short execution times of pulse-based approaches~\cite{Meitei2021,Asthana2023,Egger2023,Meirom2023,Sherbert2025,Long2024MinEvolTimes}?

In this Article, we give a positive answer to this question.
Leading VQE algorithms have circuits composed of a family of elements called qubit excitations, which mimic spin-orbital transitions~\cite{Yordanov_2020, Yordanov_2021, Anastasiou2024, Long24}. The implementation of these circuit elements requires $10$ or $34$ primitive gates. We construct a pulse-based methodology for their direct implementation. To showcase our methodology, we use the specific example of a silicon spin-qubit quantum processor. In simulations, we show that our pulse-optimisation package reduces the implementation time of qubit-excitation elements by up to \improvementUpTo~times compared with the gate-based approach (Tab.~\ref{tab:mainresultsconcise}). 
Thus, we cut the temporal cost of implementing the fundamental components of leading VQE algorithms. We emphasise that this speed-up, which naturally will improve algorithmic noise resilience, is scalable---it does not depend on problem sizes.  Moreover, we explore aspects of pulse-based qubit excitations relevant to practical implementation. To simulate arbitrary molecular wavefunction transitions, qubit-excitation elements must have tunable parameters, called \textit{excitation strengths}. Our pulses' temporal profiles change continuously with the excitation strength. Thus, it is sufficient to pre-optimise pulses for a finite number of excitation strengths---high-fidelity pulses at intermediate excitation strength can be obtained via interpolation. Moreover, the pulses have simple forms. They rely on exchange couplings only; no magnetic microwave drives are required to simulate qubit-excitation elements.

The remainder of this article is structured as follows.\ \cref{sec: results} begins with an brief description of VQE algorithms, qubit-excitation elements and the silicon spin-qubit device Hamiltonian. Then, we describe how we composed our benchmark runtimes of conventional gate-based approaches for implementing qubit excitations.  Next, we provide our main results regarding the implementation times and operational features of pulse-optimised qubit excitations on silicon hardware. Finally, in \cref{sec: discussions}, we conclude with a discussion of our findings. 

\section{Results}\label{sec: results}
\begin{figure}
    \centering
    \includegraphics[width=\columnwidth]{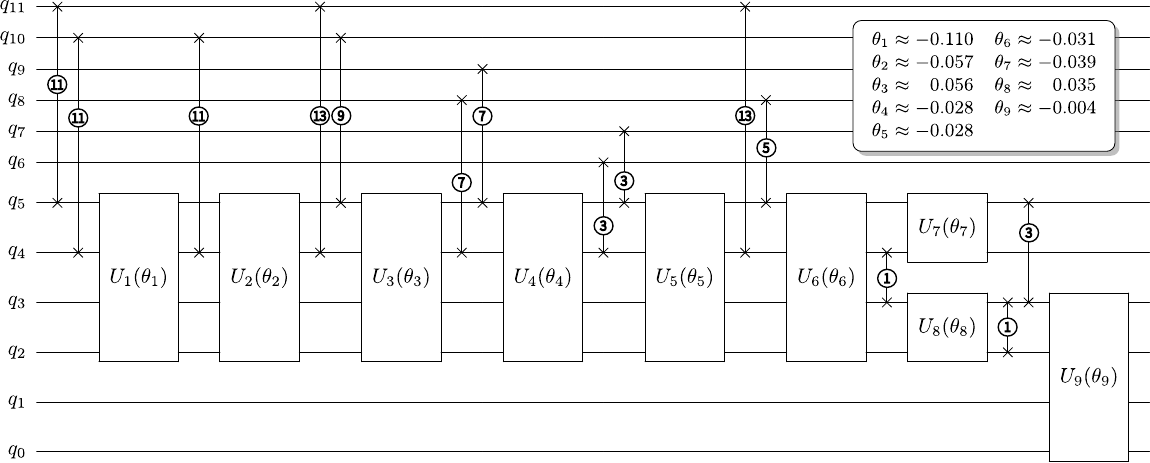}
    \caption{
    \textbf{VQE circuit to prepare an estimate of the LiH ground state.} Starting from a 12-qubit Hartree--Fock state, the circuit uses SWAP operations and single- and double-qubit excitations [Eqs.~\eqref{eq:defSingleQubitExcitation} and~\eqref{eq:defDoubleQubitExcitation}] to approximate the groundstate of LiH. A classical optimiser is used to fine-tune the excitation strength ($\theta_i$) of the circuit elements. The circled numbers on the SWAP operations indicate the number of nearest-neighbour SWAP operations required in a linear architecture. The circuit was constructed using numerical simulations of the QEB-ADAPT-VQE algorithm~\cite{Yordanov_2021}.
    }\label{fig:gateDecompositionLiH}
\end{figure}

\textit{Qubit excitations:---}VQE algorithms for chemistry applications rely on parameterised \textit{ansatz} circuits $U(\vec{\theta})$ to prepare eigenstate approximations of molecular Hamiltonians.
The main difference between VQEs is in the construction of $U(\vec{\theta})$. Normally, $U(\vec{\theta}) = U_N(\theta_N) \cdots U_1 (\theta_1)$ is decomposed in terms of $N$ circuit elements $U_i$, each parameterised by a single parameter $\theta_i$. See Fig.~\ref{fig:gateDecompositionLiH} for an example circuit.  Some VQEs use fixed sequences of circuit elements~\cite{peruzzo2014originalVQE,Romero_2019,Lee2019}, whilst others grow molecule-tailored circuits adaptively throughout the algorithms~\cite{GrimsleyAdapt,Tang2021,Yordanov_2021,Zhang2021,Gomes2021,Yordanov2022eQEBAdapt,OverlapAdaptVQE_2023,Shkolnikov2023avoidingsymmetry,Fitzpatrick2024,Anastasiou2024,Long24,Ramoa2025}. In addition, the circuit elements themselves differ between VQEs. Hardware-efficient circuit elements are designed to reduce the number of CNOT gates in VQE circuits~\cite{peruzzo2014originalVQE,Ryabinkin2018,Tang2021}. Physically-motivated circuit elements incorporate the physics of electronic spin orbitals in the VQE circuits at the expense of additional gates~\cite{Romero_2019,Lee2019,Yordanov_2021,Ramoa2025,Yordanov_2020}. 

We focus on qubit-excitation elements, which combine hardware efficiency with physical motivation~\cite{Yordanov_2020,Yordanov_2021}. Qubit-excitation elements mimic single- or double-electron hopping between spin orbitals, but do not explicitly track the fermionic phase, which arises through the anti-commutation relations. Instead, the phase is corrected by the classical optimiser. There are strong heuristic indications that adaptive VQE algorithms based on qubit-excitation elements can lead to scalable approaches to calculate properties of mid-size molecules~\cite{Yordanov_2021,Grimsley2023}. Therefore, we focus our study on these circuit elements. However, the methodology developed below is general and could also be applied to other circuit elements. 

To simulate spin orbitals on a quantum computer, one requires a map to translate the fermionic Hilbert space to the qubit Hilbert space.
We use the Jordan--Wigner encoding~\cite{jordanwigner}\nocite{Duck1997}. In this encoding, each qubit 
$\ket{q}_i$ represents the $i$th electronic spin orbital. $\ket{1}_i$ and  $\ket{0}_i$ denote an occupied and unoccupied $i$th orbital, respectively. Commonly, these orbitals are chosen from a standardised basis set such as the STO-3G basis set~\cite{ditchfield1971self}. 

\begin{figure*}[tb]
    \centering
    \begin{tabular}{ccc}
        \scalebox{0.68}{\begin{quantikz}[column sep=0.2cm, row sep=0.5cm]
            \lstick{$q_k$} & \ghost{R_z\left(\frac{\pi}{2}\right)} & \gate[wires=2]{A_{ik} (\theta)} & \qw \\
            \lstick{$q_i$} & \ghost{R_z\left(\frac{\pi}{2}\right)} &                  & \qw
        \end{quantikz}}  
        & $\cong$ & 
        \scalebox{0.68}{\begin{quantikz}[column sep=0.3cm, row sep=0.5cm]
        \lstick{$q_k$} & \gate{R_z (\frac{\pi}{2})} & \gate{R_x (\frac{\pi}{2})} & \ctrl{1} & \gate{R_x (\theta)}  & \ctrl{1} & \gate{R_x (-\frac{\pi}{2})} & \gate{R_z (-\frac{\pi}{2})} & \qw  \\
        \lstick{$q_i$} & & \gate{R_x (\frac{\pi}{2})}& \targ{} & \gate{R_z (\theta)}      & \targ{} & \gate{R_x (-\frac{\pi}{2})} & \qw & \qw
    \end{quantikz}} \\
    \vspace{0.1cm} & & \\
    \scalebox{0.68}{
        \begin{quantikz}[column sep=0.2cm, row sep=0.5cm]
            \lstick{$q_l$} & \ghost{R_y\left(\frac{\theta}{8}\right)} & \gate[wires=4]{A_{ijkl} (\theta)} & \qw \\
            \lstick{$q_k$} & \ghost{X}                            &                                  & \qw \\
            \lstick{$q_j$} & \ghost{R_y\left(\frac{\theta}{8}\right)} &                                  & \qw \\
            \lstick{$q_i$} & \ghost{X}                            &                                  & \qw
        \end{quantikz}}
    & $\cong$ & 
    \scalebox{0.68}{
\begin{quantikz}[column sep=0.1cm, row sep=0.5cm]
\lstick{$q_l$} & \ctrl{1} & \qw      & \ctrl{2} & \gate{R_y\left(\frac{\theta}{8}\right)} & \ctrl{1} & \gate{R_y\left(-\frac{\theta}{8}\right)} & \ctrl{3} & \gate{R_y\left(\frac{\theta}{8}\right)} & \ctrl{1} & \gate{R_y\left(-\frac{\theta}{8}\right)} & \ctrl{2} & \gate{R_y\left(\frac{\theta}{8}\right)} & \ctrl{1} & \gate{R_y\left(-\frac{\theta}{8}\right)}& \ctrl{3} & \gate{R_y\left(\frac{\theta}{8}\right)} & \ctrl{1} & \gate{R_y\left(-\frac{\theta}{8}\right)} & \ctrl{2} & \gate{R_z\left(\frac{\pi}{2}\right)}  & \qw                                   & \ctrl{1} & \qw  \\
\lstick{$q_k$} & \targ{}  & \gate{X} & \qw      & \gate{H}                                & \targ{}  & \qw                                      & \qw      & \qw                                     & \targ{}  & \qw                                      & \qw      & \qw                                     & \targ{}  & \qw                                     &  \qw     & \qw                                     & \targ{}  & \gate{H}                                 & \qw      & \qw                                   & \gate{X}                              & \targ{}  & \qw  \\
\lstick{$q_j$} & \ctrl{1} & \qw      & \targ{}  & \qw                                     & \qw      & \qw                                      & \qw      & \qw                                     & \qw      & \gate{H}                                 & \targ{}  & \qw                                     & \qw      & \qw                                     &  \qw     & \qw                                     & \qw      & \gate{R_z\left(-\frac{\pi}{2}\right)}    & \targ{}  & \gate{R_z\left(-\frac{\pi}{2}\right)} & \gate{R_y\left(-\frac{\pi}{2}\right)} & \ctrl{1} & \qw  \\
\lstick{$q_i$} & \targ{}  & \gate{X} & \qw      & \qw                                     & \qw      & \gate{H}                                 & \targ{}  & \qw                                     & \qw      & \qw                                      & \qw      & \qw                                     & \qw      & \qw                                     &  \targ{} & \gate{H}                                & \qw      & \qw                                      & \qw      & \qw                                   & \gate{X}                              & \targ{}  & \qw      
\end{quantikz}} \\
    \end{tabular}
    \caption{\textbf{Decomposition of qubit-excitation elements into primitive gates.} The top panel displays the circuit for a single-qubit excitation; the bottom panel displays the circuit for a double-qubit excitation~\cite{Yordanov_2021}. The qubit-excitation elements require 10 and 34 gates, respectively. 
             }\label{fig:CircuitDiagrams} 
\end{figure*}
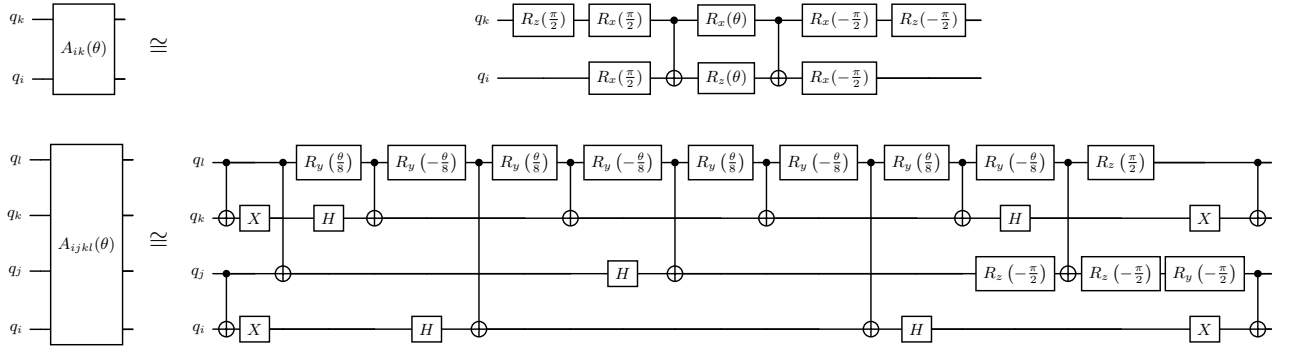%
We can define the qubit-excitation elements within the Jordan--Wigner encoding. First, we introduce the \textit{qubit creation} and \textit{annihilation operators}:
\begin{align}
    Q_i^\dagger &= \frac{1}{2}(X_i - i Y_i) \quad\text{and} \\
    Q_i &= \frac{1}{2}(X_i + i Y_i) .
\end{align}
Exponentiating combinations of these operators yields the single-qubit and double-qubit excitation elements:
\begin{align}
    \label{eq:defSingleQubitExcitation}
    A_{ik}(\theta) &\equiv \exp[\theta(Q_i^\dagger Q_k - Q_k^\dagger Q_i)]\quad\text{and} \\
    A_{ijkl}(\theta) &\equiv \exp[\theta(Q_i^\dagger Q_j^\dagger Q_k Q_l - Q_k^\dagger Q_l^\dagger Q_i Q_j)].
    \label{eq:defDoubleQubitExcitation}
\end{align}
The parameter $\theta$ is the excitation strength of the circuit element and is subject to optimisation in VQEs.
In the computational basis, the matrix representation of a single-qubit excitation between qubit $i$ and qubit $k$ is
\begin{align}
    \label{eq:SQEmatrix}
    A_{ik}(\theta) \;  \mapsto \;
    \begin{pmatrix}
        1&0&0&0 \\
        0&\cos(\theta)&-\sin(\theta)&0 \\
        0&\sin(\theta)&\cos(\theta)&0 \\
        0&0&0&1 \\
    \end{pmatrix} .
\end{align}
$A_{ik}(\theta)$ rotates the subspace spanned by $\ket{01}$ and $\ket{10}$, keeping the remaining space fixed. 
Similarly, the double-qubit excitation between qubits (1,2) and qubits (3,4) acts only as a rotation between $\ket{0011}$ and $\ket{1100}$. 
Although their matrix representations are relatively simple, a large number of primitive gates are needed to generate the quantum circuits of qubit excitations  (see Fig.~\ref{fig:CircuitDiagrams})~\cite{Yordanov_2020}. 
Below, we abandon the gate-based model of quantum computing and construct pulses that simulate the qubit excitations in a shorter time.

\textit{Silicon quantum processor:---}A pulse-based algorithm must be tailored to a specific hardware. To showcase our methodology, we focus on electron spin qubits in semiconductor quantum dots~\cite{LossDiVincenco1998,Zwanenburg2013,Burkard2023}. Our choice is motivated by silicon quantum dots' compatibility with industrial foundries and the qubits' small size~\cite{Zwerver2022Mar,Elsayed2023,Li2020,Elsayed2024Jul,Neyens2024,Koch2025}. Additionally, there are impressive experimentally demonstrated features of electron spin qubits.  Single- and two-qubit fidelities already exceed 99\%~\cite{Steinacker2025Oct, Huang2024Mar, Tanttu2024Nov,Xue2022Jan,Philips2022,Noiri2022,wu2025simultaneoushighfidelitysinglequbitgates}, coherence times can be as high as $T_2^* \approx 100~\micro$s~\cite{Stano2022}, and a $6$-qubit device has been demonstrated~\cite{Philips2022}. These qualities make electron-spin systems attractive candidates for realising quantum computers~\cite{2026Peter}.  While we focus on electron spin qubits, our pulse-based acceleration of qubit excitations is readily adaptable to other hardware.

\begin{figure}
    \centering
    \includegraphics[width=\linewidth]{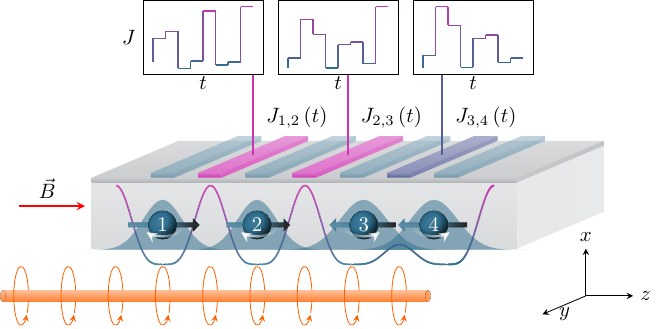}
    \caption{\textbf{Schematic of a silicon MOS device.} 
Single-electron spin-qubits (arrows) are confined in Si (light grey) by potential wells (solid curve) created by plunger (semitransparent blue) and barrier gates, also called $J$-gates (purple/pink) on top of an SiO$_x$ layer (dark grey). The $J$-gates control the potential barriers between electrons, and thus, the exchange interactions. The electrons' spatial wavefunctions are shown as blue shaded bell curves.  Each electron spin is subject to a static, external, magnetic field. A time-dependent magnetic field can be generated by a microwave antenna (orange). The white arrows illustrate the action of the double-qubit excitation.
}\label{fig:SiMOSdevice}
\end{figure}

The specific quantum-dot platform we model is composed of Loss--DiVincenzo electron spin-qubits~\cite{LossDiVincenco1998} in a linear array (see Fig.~\ref{fig:SiMOSdevice}).  In such a device, voltages on metal gates confine electron spins to a silicon layer in a heterostructure. 
We denote by $\vec \sigma^{(i)} = (X_i,Y_i,Z_i)$ a vector of Pauli matrices acting on the $i$th spin.
An external magnetic field induces a Zeeman Hamiltonian $B_i Z_i/2$ on each spin. The qubit is defined by the down and up states of the spin in the co-rotating frame of this Zeeman term. 
Single-qubit rotations can be implemented by applying a microwave drive, $g(t)=\Omega(t)\cos(\omega t-\varphi)$, (either via an antenna~\cite{Dehollain_2013, Huang2024Mar, Tanttu2024Nov, Steinacker2025Oct} or---in the presence of a micromagnet---via a confinement gate~\cite{Xue2022Jan,Noiri2022,Philips2022}). Two-qubit control is achieved by varying the voltage levels on barrier gates. These voltages control the exchange interactions $J_{i,j}(t)$ between the $i$th and $(i+1)$th spins~\cite{Burkard2023}. The quantum dynamics of the spin qubits in our device are generated by a Heisenberg chain~\cite{Burkard2023}:
\begin{align}
    \label{eq:deviceHamiltonian}
    H=-\sum_{i=1}^N\frac{B_i}{2}Z_i + g(t) \sum_{i=1}^N X_i 
    +\sum_{i=1}^{N-1}\frac{J_{i,i+1}(t)}{4} \vec \sigma^{(i)} \vec \sigma^{(i+1)}.
\end{align}
Typical device parameters are summarised in Tab.~\ref{tab:device_parameters}. These parameters apply to both foundry-compatible silicon devices~\cite{Steinacker2025Oct, Tanttu2024Nov, Huang2024Mar} and silicon-germanium heterostructures~\cite{Philips2022, Xue2022Jan, Noiri2022,wu2025simultaneoushighfidelitysinglequbitgates}.
\begin{table}[tb]
\renewcommand{\arraystretch}{1.8}
\centering
\begin{ruledtabular}
\begin{tabular}{ll}
\textbf{Parameter} & \parbox{0.62\linewidth}{\textbf{Value}}\\
\colrule
Exchange &
\parbox[t]{0.62\linewidth}{$J_{i,i+1}/h \leq 10\,\mathrm{MHz}$~\cite{Tanttu2024Nov,Philips2022}} \\
Zeeman &
\parbox[t]{0.62\linewidth}{$\bar{B}_i/h \lesssim 18\,\mathrm{GHz}$} \\
Detuning &
\parbox[t]{0.62\linewidth}{
$(B_{i+1}-B_i)/h = 8\,\mathrm{MHz}$ (without micromagnet)~\cite{Huang2024Mar}\\
$\leq 300\,\mathrm{MHz}$ (with micromagnet)~\cite{Philips2022}
} \\
Microwave &
\parbox[t]{0.62\linewidth}{$\Omega/h \leq 5\,\mathrm{MHz}$~\cite{Huang2024Mar,Philips2022}}
\end{tabular}
\end{ruledtabular}
\caption{\textbf{Typical device parameters for electron spin qubits.}
The detuning depends on whether a micromagnet is absent or present.
}\label{tab:device_parameters}
\end{table}

\textit{Gate-based qubit excitations:---}To provide a benchmark for our results, we first establish the gate-based execution times of single- and double-qubit excitations on silicon processors. A detailed analysis is provided in the Supplementary Material, which we briefly summarise here.
High‑fidelity qubit operations for Loss–-DiVincenzo spin qubits typically employ the following native gates: (i) a controlled-$Z$ (CZ) gate~\cite{Huang2024Mar, Tanttu2024Nov, Steinacker2025Oct, Xue2022Jan, Philips2022, Rimbach-Russ2023Sep}, (ii) an $\Xpih$ gate together with virtual $\VZ$ gates to implement arbitrary single‑qubit rotations~\cite{McKay17}, and (iii) a resonant SWAP  gate~\cite{Rimbach-Russ2023Sep}. To implement the qubit excitations shown in Fig.~\ref{fig:CircuitDiagrams}, we replace each CNOT gate by a CZ  gate conjugated by two Hadamard ($\text{H}$) gates. Consequently, implementing a single- or double-qubit excitation requires executing 2 or 11 layers of CZ gates, respectively. Assuming that the CZ gates within each layer can be executed in parallel, each layer incurs the execution time $T_{\text{CZ}}$ of a single CZ gate.
Further, any arbitrary single-qubit rotation can be decomposed into three virtual $\VZ$ gates and two $\Xpih$ gates~\cite{McKay17}. Since the $\VZ$ gates are virtual, they incur no execution time. Thus, implementing a single- or double‑qubit excitation additionally requires executing $3$ or $12$ layers of two consecutive $\Xpih$ gates, respectively. Assuming that the $\Xpih$ gates in each layer can be executed in parallel, each such layer incurs the execution time $T_{\Xpih}$.
Assuming that the double-qubit excitation is executed on a linear device, \textit{i.e.~}a one-dimensional spin chain with nearest neighbour interaction, six additional SWAP gates with execution time $T_{\text{SWAP}}$ are required.
Thus, the execution times of single- and double-qubit excitations on a linear device are
\begin{align}
    T_{\text{SQ}} 
    & = 2 \cdot T_{\text{CZ}} + 3 \cdot 2 \cdot T_{\Xpih}, \\
    T_{\text{DQ}}
    & = 11 \cdot T_{\text{CZ}} + 12 \cdot 2 \cdot T_{\Xpih} + 6 \cdot T_{\text{SWAP}}.
\end{align}
Finally, assuming that all gates are implemented using state-of-the-art pulse-shaping techniques for high‑fidelity operations~\cite{Rimbach-Russ2023Sep}, we derive lower bounds on the execution times $T_{\Xpih}$, $T_{\text{CZ}}$, and $T_{\text{SWAP}}$ as functions of the detuning. For a device with $J_{\max}/h=10$~MHz, $\Omega_{\max}/h=5$~MHz, and $\Delta/h\in[2,300]$~MHz, we find the execution times reported in Tab.~\ref{tab:gate_times}. We note that particularly long execution times arise at small detunings, for devices which do not have a micromagnet. Further details are provided in the Supplementary Material.
\begin{table}[tb]
\renewcommand{\arraystretch}{1.33}
\centering
\begin{ruledtabular}
\begin{tabular}{ccccc}
Detuning &
\multicolumn{2}{c}{with micromagnet} &
\multicolumn{2}{c}{no micromagnet} \\
$\Delta/h$ (MHz) &
300 &
100 &
8 &
2 \\ \colrule
$T_{\Xpih}$ (ns) &
$\geq 56$ &
$\geq 100$ &
$\geq 259$ &
$\geq 453$ \\
$T_{\mathrm{CZ}}$ (ns) &
$\geq 55$ &
$\geq 98$ &
$\geq 196$ &
$\geq 247$ \\
$T_{\mathrm{SWAP}}$ (ns) &
$\geq 100$ &
$\geq 100$ &
$\geq 100$ &
$\geq 100$ \\ \colrule
$T_{\mathrm{SQ}}$ (ns) &
$\geq 446$ &
$\geq 796$ &
$\geq 1946$ &
$\geq 3212$ \\
$T_{\mathrm{DQ}}$ (ns) &
$\geq 2549$ &
$\geq 4078$ &
$\geq 8972$ &
$\geq 14189$
\end{tabular}
\end{ruledtabular}
\caption{\textbf{Typical execution times of essential native gates and qubit excitations.}
The data assumes a device with $J_{\max}/h=10\,\mathrm{MHz}$, $\Omega_{\max}/h=5\,\mathrm{MHz}$, and $\Delta/h\in[2,300]\,\mathrm{MHz}$.
The execution times at $\Delta/h=300,100,8$, and $2\,\mathrm{MHz}$ use a Tukey, a Hann, a Kaiser, and a Kaiser window, respectively.
For further details, see the Supplementary Material.
}\label{tab:gate_times}
\end{table}
In summary, we find that a gate-based approach would require \TimeSQEbenchmark~ns or \TimeDQEbenchmark~ns
to implement a single- or double-qubit excitations, respectively. Consequently, one can implement only \fpeval{floor (100000 / \TimeDQEbenchmarkHigh)}--\fpeval{floor (100000 / \TimeDQEbenchmarkLow)} double-qubit excitations during the typical coherence time of current devices ($T_2^*\lesssim100\ \micro$s). 
This poses a problem as chemically-accurate simulations of small molecules such as BeH$_2$ and H$_6$ require hundreds of qubit excitations~\cite{Yordanov_2021}.

\textit{Qubit excitations via optimised pulses:---}We now demonstrate how one can significantly shorten the execution times of qubit excitations by switching from the standard gate-based approach to a pulse-based implementation. To do so, we expand the pulse-shaping method from Ref.~\cite{Long2024MinEvolTimes}.  (For details, see the Methods.)
For a pulse of duration $T$, we numerically integrate the device Hamiltonian [Eq.~\eqref{eq:deviceHamiltonian}] for a specific pulse $\{J_{i,i+1}(t), g(t)\}$. This yields a corresponding unitary $U(T)$. We then use the gradient-ascent pulse-engineered (GRAPE) algorithm~\cite{Khaneja2005Feb} to shape the pulse such that $U(T)$ mimics a target qubit excitation $A(\theta)$ [Eqs.\eqref{eq:defSingleQubitExcitation} and~\eqref{eq:defDoubleQubitExcitation}]. We optimise the pulses to minimise the average gate infidelity~\cite{NIELSEN2002249}
\begin{align}
\label{Eq:InFid}
    \mathcal{I}(U(T),A(\theta)) = 1 - \frac{\left|\mathrm{Tr}(U{(T)}^{\dagger} A(\theta))\right|^2 + d}{d(d+1)}.
\end{align}
Here $d$ is the dimensionality of the Hilbert space on which $U(T)$ and $A(\theta)$ act non-trivially. The smaller $\mathcal{I}(U(T),A(\theta))$, the better the unitary approximation.

An important observation of our numerical simulations is that $U(T)$ can approximate well any $A(\theta)$ in the absence of microwaves.
Hence, all our numerical simulations present data where $g(t) =0 $. The exchange-only pulses $J_{i,i+1}(t)$ are composed of piecewise constant segments. We find that high-fidelity pulses are possible with segment durations longer than two nanoseconds. This is compatible with current hardware~\cite{Rotta2017Jun_segmentwidth}. 

A key quantity in our simulations is the minimal pulse duration at which $U(T)$ can mimic $A(\theta)$ with infidelity smaller than $\epsilon$. We call this quantity the minimal evolution time (MET), and calculate it as in~\cite{Deffner2017}:
\begin{align}
    \mathrm{MET}:=\min\left\{T>0 \; : \; \mathcal{I}(U (T),A(\theta))<\epsilon\right\}.
\end{align}
We find that single-qubit excitations can be implemented \improvementFactorSQE~times faster using our pulses than with the gate-based approach. The gain for the double-qubit excitation is even larger; they can be implemented \improvementFactorDQE~times faster. 
Our results are summarised in Tab.~\ref{tab:mainresultsconcise}. 

\begin{figure*}[ht]
    \centering
    \includegraphics[width=\textwidth]{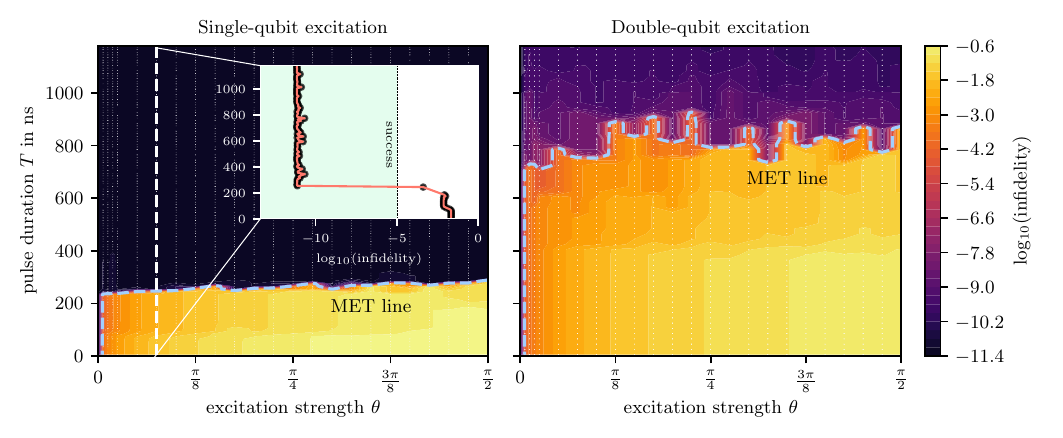} 
    \caption{ \textbf{Performance of pulse-based qubit excitations.}
        Infidelity of a single- (left) and double- (right) qubit excitation plotted as a function of excitation strength $\theta$ and pulse duration $T$. Dark regions indicate $(\theta, T)$-combinations for which we found low-infidelity pulses that approximate a qubit-excitation well. Lighter colours indicate regions of high infidelity, where we were unable to find pulses that approximate qubit excitations. 
        The blue dashed line (where the infidelity equals $10^{-5}$) marks the transition between these regions.  The inset highlights a vertical cut through our data.  Each black dot represents the outcome of the most successful pulse simulation at a fixed $T$. At large $T$, we find high-fidelity pulses. Below a minimal evolution time, the optimisation algorithm is not able to shape a pulse to approximate the target qubit excitation with high fidelity. The device parameters used for these results correspond to the left column of Tab.~\ref{tab:device_parameters}.
    }\label{fig:thetaTmap}
\end{figure*}

Next, we investigate how the MET depends on the excitation strength. 
In Fig.~\ref{fig:thetaTmap}, we provide a detailed analysis of how the pulse duration and the circuit elements' excitation strength $\theta$ affect the maximum attainable fidelity (minimum attainable infidelity).
The contours in Fig.~\ref{fig:thetaTmap} show the pulse-induced evolutions' infidelities with respect to a single- (left) and double- (right) qubit excitation. The darker the colour, the lower the infidelity. 
The excitation strength $\theta$ changes with the horizontal axis, and the pulse duration $T$ changes with the vertical. For long durations $T$, our pulse-construction framework is more likely to find a pulse that approximates the target circuit element with low infidelity. 
Therefore, to produce the data in the figure, we initialise an ensemble of randomly generated pulses with a long duration $T\approx 1200$~ns. Keeping $T$ fixed, we optimise the pulses in the ensemble and plot the data from the most successful (lowest infidelity) pulse. To obtain shorter pulses (moving vertically downward in the figure), we initialise our optimiser with a compressed version of the longer pulses and iteratively reduce $T$. (See the Methods for details.) We plot the data from the best-case pulses.
Figure~\ref{fig:thetaTmap} displays a sharp transition from high- to low-infidelity pulses above a specific pulse duration $T$ (see the inset). 
The $T$-value at this transition is the MET\@.
We highlight the MET as a function of $\theta$ with a blue dashed curve. For pulse durations shorter than the MET, there are no pulses with infidelities $\mathcal{I}$ below $\epsilon = 10^{-5}$. The METs, we find, follow two trends. First, the MET is $T \approx 0$ for $\theta \approx 0$. This is expected as qubit excitations equal the identity operation at $\theta=0$. Second, the MET is remarkably flat with respect to $\theta$ at $T \approx 250$~ns and $T \approx 820$~ns for the single- and double-qubit excitations, respectively. The spin-qubit processor's control fields have to act for approximately the same amount of time, irrespective of the excitation strength $\theta$. 

Whilst our qubit-excitation pulses are \improvementFactorSQE~and \improvementFactorDQE~times faster than their gate-based equivalents, they are significantly slower than some of the pulse-based molecular-state-preparation times presented in previous work~\cite{Long2024MinEvolTimes}. In state preparation, one typically evolves a \textit{fixed} and known initial state towards a target final state. This enables the construction of a faster pulse towards the target state.
In our work, instead, we design an optimised pulse that induces a desired evolution on \textit{any} input state.  Our approach combines the temporal efficiency of pulse-based methods with the scalability of modular VQEs, that are composed of qubit-excitation elements.

\begin{figure*}[tb]
        \centering
        \includegraphics[width=\textwidth]{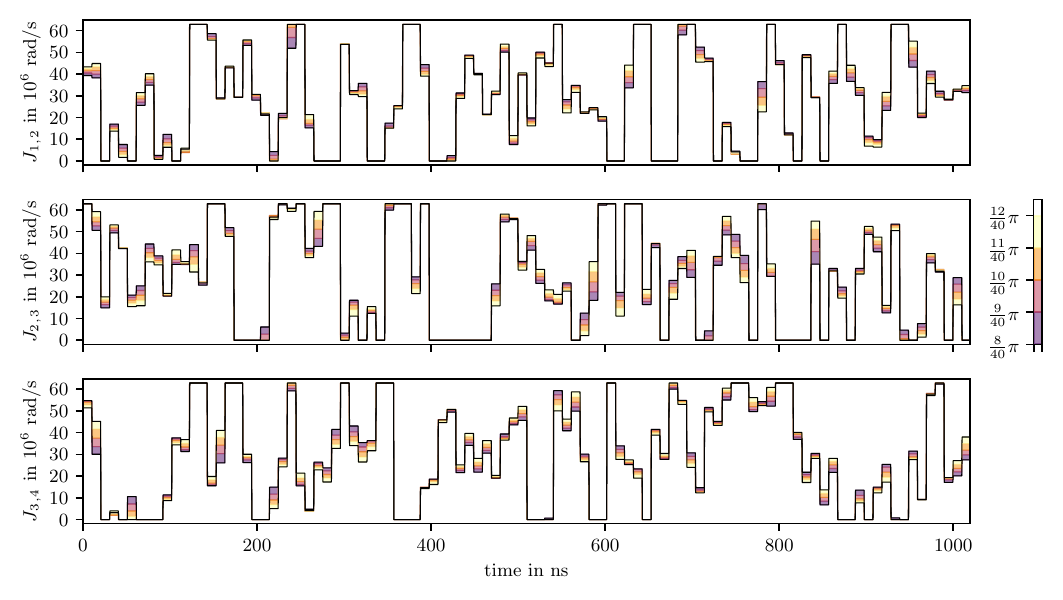}
        \caption{ \textbf{Pulse shape vs pulse strength.}
        The optimised pulse shapes of double-qubit excitations are plotted for five excitation strengths $\theta$ in the interval $[\frac{8}{40}\pi, \frac{12}{40}\pi]$. 
        The black lines mark the pulses corresponding to the two interval boundaries. The coloured rectangles mark the difference in segment amplitude for the different $\theta$ values. Large and small coloured rectangles indicate fast and slow changes in segment height with $\theta$, respectively. The absence of coloured rectangles implies that the corresponding segment amplitude is constant across the $\theta$ range.
        }\label{fig:continuouspulses}
    \end{figure*}

\textit{$\theta$-dependence and pulse interpolation:---}For practical purposes, it is important to investigate how our optimised pulses vary under small changes in the qubit-excitation strength $\theta$. 
As $\theta \in [0,\pi/2]$ is a continuous parameter, there are an infinite number of qubit-excitation elements. Naturally, one cannot optimise an infinite number of pulses. Instead, any experimental procedure must rely on many pre-optimised pulses and interpolation to meet the needs of a specific algorithm's implementation. We now present evidence that indicate that such pulse interpolation can be effective. First, we analyse the $\theta$-dependence of a specific pulse sequence for a double-qubit excitation. Figure~\ref{fig:continuouspulses} shows the $J$-drive amplitudes as a function of time. 
Whilst the pulses have seemingly random amplitude profiles for a fixed value of $\theta$, they vary continuously with small $\theta$ changes.
This continuous behaviour indicates that one can interpolate between pulses optimised for two nearby values $\theta_1$ and $\theta_2$ to approximate a pulse corresponding to an intermediate strength $ \theta^{\prime}$. 
\begin{figure}
    \centering
    \includegraphics[width=\linewidth]{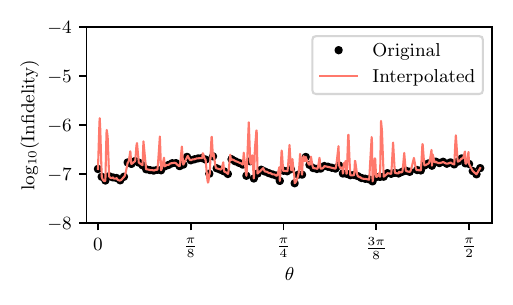}
    \caption{ \textbf{Performance of interpolated double-qubit-excitation pulses.}
   The black dots represent the baseline data for linear pulse interpolation: each black dot marks the infidelity of a pulse optimised specifically for a distinct value of $\theta$. The red curve shows the infidelity achieved for an interpolated pulse at an intermediate value of $\theta$. Throughout this figure, the pulse duration is $T=1020$~ns.
   }\label{fig:interpolationquality}
\end{figure}
Next, we benchmark the performance of pulse interpolation.  We pick $100$ evenly spaced values of $\theta \in [0,\pi/2]$. For each value of $\theta$, we optimise a pulse to implement double-qubit excitations to an infidelity of $\mathcal{I}\approx 10^{-7}$.  The infidelities of the $100$ baseline pulses are plotted as black dots in Fig.~\ref{fig:interpolationquality}. 
To obtain pulses for double-qubit excitations with $\theta$-values between the black dots, we use interpolation. For each $\theta$ between pre-optimised $\theta_1$ and $\theta_2$, we linearly interpolate between the pulse amplitudes of the $\theta_1$- and $\theta_2$-pulse (see Methods for more detail).
The red curve tracks the infidelities of the interpolated pulses. In most cases, the interpolated pulses achieve infidelities that match the baseline data. Between some points, the infidelity reaches $10^{-6}$, which still constitutes a minor error. We therefore conclude that a simple linear-interpolation routine is sufficient to provide pulses for the entire $\theta$-interval. 

\begin{figure}[tb]
    \centering
    \includegraphics[width=\linewidth]{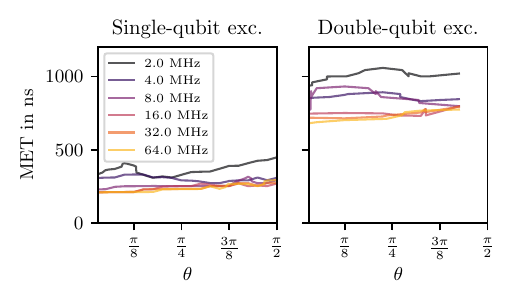}
    \caption{ \textbf{METs for different detunings. }
    The single- (left) and double- (right) qubit-excitation METs for different magnetic detunings are plotted as functions of the excitation strength $\theta$. The colours indicate different detuning levels.
    }\label{fig:metlinesperdetuning}
\end{figure}

\textit{METs as a function of detuning:---}We now investigate our methodology's stability to hardware variations in the Zeeman-energy detunings. 
Our results, so far, were generated from numerical simulations with detunings set to $8$~MHz. This is indicative of spin-qubit devices without micromagnets~\cite{Huang2024Mar}. However, when utilising micromagnets, detuning values as high as $630$~MHz between neighbouring qubits have been reported~\cite{Yoneda2023}.
To broaden the scope of our numerical investigation, we calculated the qubit-excitations' METs for a range of detunings. The results are shown in Fig.~\ref{fig:metlinesperdetuning}. 
The detuning has little effect on the MET line's characteristic: It remains relatively flat w.r.t. $\theta$. In general, the METs decrease slowly with increasing detunings. 
Larger detunings, therefore, facilitate faster qubit-excitation implementations. However, as shown in Fig.~\ref{fig:metlinesperdetuning}, this effect saturates: qubit excitations cannot be made arbitrarily fast by increasing the neighbour detuning. The overall sensitivity of the MET-line to the detuning is rather weak. In summary, our methodology for producing MET pulses for qubit excitation is robust to variations in device parameters. 
\textit{Qubit excitations over idling qubits:---} Most current electron spin-qubit devices are one-dimensional spin chains with limited connectivity~\cite{Burkard2023}. Our simulations have only considered single- and double-qubit excitations between neighbouring spin qubits and neighbouring pairs of spin qubits, respectively. However, general quantum chemistry algorithms will naturally require the simulation of circuit elements between non-neighbouring qubits. Non-local single- and double-fermionic or qubit excitations can be achieved by conjugating our single- and double-qubit excitations pulses with a CNOT staircase culminating in a CZ gate or a SWAP staircase, respectively~\cite{Yordanov_2020}.

\section{Discussion}\label{sec: discussions}

Hybrid quantum-classical algorithms based on VQEs form an appealing route to near-term and useful quantum-chemistry calculations. Previous works have highlighted the benefits and disadvantages of gate-based and pulse-based VQEs. The iteratively grown problem-tailored gate circuits of the ADAPT-VQE algorithms are believed to generate reasonable sampling and trainability overheads for larger molecular systems~\cite{GrimsleyAdapt, Yordanov_2021, Grimsley2023}. However, these circuits incur fatal decoherence during their relatively long runtimes~\cite{Dalton2024, Long24}. Pulse-based methods enable the implementation of small-molecule VQE circuits within the coherence times of current hardware~\cite{Meitei2021, Asthana2023, Egger2023, Meirom2023, Sherbert2025, Long2024MinEvolTimes}. However, these methods incur intractable sampling and optimisation overheads for larger molecules~\cite{Sherbert2025, Long2024MinEvolTimes}. Our work combines the runtime advantage of pulse-based approaches with the scalability of adaptive VQE algorithms. 

The ADAPT-VQE algorithms construct circuits from modular elements. A popular set of such elements contains single- and double-qubit excitations~\cite{Yordanov_2021, Yordanov_2020}. Normally, the qubit excitations are implemented with 10 or 34 standardised gates. Here, we have constructed a framework to implement the qubit excitations directly, instead, using optimised pulses for a specific hardware's control fields.
We showcased our methodology by simulating a silicon electron-spin processor. We found that single-qubit excitations can be implemented in $\leq 289$~ns, reducing the runtime by a factor of \improvementFactorSQE~compared with gate-based methods. For double-qubit excitations, the reduction is even more dramatic. The runtime is reduced by a factor of \improvementFactorDQE, to $\leq$ \TimeDQEpulsebased~ns. 
Thus, we anticipate that our work has brought us yet another step closer to useful chemistry calculations on near-term quantum hardware.  We conclude this article with a discussion of five features of our pulse-based qubit excitations.

First, we focused our study on the optimisation of pulses for qubit-excitation elements implemented on electron-spin processors. However, we expect our methods to be readily applicable to other architectures and circuit elements. We leave these extensions as open problems for future work.

Second, pulse-optimised qubit excitations' infidelities display abrupt transitions when the pulse duration falls below the MET (sharp change in colour in Fig.~\ref{fig:thetaTmap}). The transition is most dramatic for the single-qubit excitations, but it is prevalent for two-qubit excitations too. This feature indicates that our numerical characterisation of METs did not suffer from local minima during the optimisation. The control fields of the SiMOS hardware simply cannot perform the operation faster.  

Third, the pulse shapes of optimised qubit excitations vary continuously with changes in the excitation strength $\theta$. From a set of $100$ pre-optimised pulses, we used interpolation to construct double-qubit excitations of arbitrary strength. The interpolated pulses simulated their target operations with infidelities of approximately $10^{-6}$. 
Thus, experimental errors are likely to be dominated by decoherence rather than optimisation inaccuracies. 

Fourth, in simulations, we found that the inclusion of single-qubit operations other than virtual $Z$ gates did not have a significant impact on the circuit-element METs.  Thus, we omitted them from our analysis. Our observation may impact the design of VQE-tailored hardware. For example, the microwave antenna used for magnetic single-qubit control is an experimental bottleneck in current devices. Our results indicate that quantum processors intended for quantum chemistry can function without such antennae.

Fifth, the qubit-excitation METs for finite excitation strengths are fairly constant. This may seem surprising as low-strength ($\theta\approx 0$) qubit excitations generate little entanglement and only weakly perturb a quantum state. Nevertheless, our pulse-optimiser cannot utilise this simplicity to cut runtimes within the constraints of the hardware. The flatness of the METs highlights a limitation of the control fields of silicon electron-spin quantum processors. It may be that other quantum-computing platforms have Hamiltonian control fields that can make better use of the functional structure of qubit excitations. We leave an investigation of this for future work.  

\section{Methods\label{sec:methods}}
\textit{Computational framework:---}We construct a numerical framework to design the pulses presented in this work. The main component of this framework is an emulator of the device Hamiltonian [Eq.~\eqref{eq:deviceHamiltonian}] and a pulse optimiser. There are several approaches to model semiconductor spin qubits~\cite{Burkard2023, ArvidssonShukur17, Lepage20}. We focus on an effective Heisenberg model [Eq.~\eqref{eq:deviceHamiltonian}] of a one-dimensional spin chain~\cite{Burkard2023}. 
Our device emulator is an extended and problem-tailored version of the one constructed in Refs.~\cite{long_2025_17116352, Long2024MinEvolTimes, long2025virtualzgatesvirtual}. It relies on the Suzuki-Trotter algorithm~\cite{Berry2005Aug, Suzuki1990Jun} to integrate the Schrödinger evolution.
We use the emulator to translate a pulse, \textit{i.e.},~time-series data for the microwave drive [$g(t)$] and exchange coupling 
[$J_{i,i+1}(t)$] into a unitary operator $U[g, J_{1,2},\ldots, J_{n-1,n}]$
on the spin qubits' Hilbert space. We label a pulse successful if this unitary has an infidelity [Eq.~\eqref{Eq:InFid}] below some threshold $\epsilon$ with respect to the target operation.  For the single-qubit excitation, we simulated the dynamics of two spin qubits, whilst for the double-qubit excitations, we simulated four spin qubits. 

We optimise the pulses to minimise their infidelity [Eq.~\eqref{Eq:InFid}] with respect to target circuit elements. To limit computational resources and to produce pulses that conform with the temporal resolution of arbitrary waveform generators, we construct the pulses using a finite number of tunable parameters. We do so using GRAPE~\cite{Khaneja2005Feb}, which naturally satisfies the temporal-resolution constraints by utilising piecewise-constant pulses. Waveforms with segment widths of 100~ps to 500~ps and rise times as low as 80~ps are possible with current hardware~\cite{Rotta2017Jun_segmentwidth}. We use $100$ segments of equal duration. Throughout our study, the pulse segments of the METs are well above 2~ns in duration. Thus, they are compatible with current hardware. 
Fig.~\ref{fig:continuouspulses} outlines how our optimiser tunes the amplitudes of the MET pulses between five different qubit excitation strengths $\theta$.

Using the SciPy implementation of the Broyden–Fletcher–Goldfarb–Shanno algorithm~\cite{Shanno1970, Fletcher2000May, Zhu1997Dec}, we optimise the amplitudes of the segments until the desired infidelity with the target unitary evolution is achieved. The single-qubit excitation and the double-qubit excitation require optimisation over 100 and 300 amplitudes, respectively. The optimisation process involves thousands of evaluations of the device emulator and of the infidelity function [Eq.~\eqref{Eq:InFid}].
Our device emulator employs QuGrad~\cite{long_2025_17393503, Long2024MinEvolTimes} to analytically compute the gradients of the infidelities with respect to the segment amplitudes. These analytics generate a significant acceleration of the optimisation procedure.

\textit{Data generation:---}The first step in our pulse design is the initialisation of the GRAPE algorithm with an ensemble of pulses consisting of 100 segments per drive channel with random amplitudes. Then, the optimiser fine-tunes these pulses, adjusting amplitudes to generate low infidelities with the target circuit element. For large pulse durations $T$, it is relatively easy to find low-infidelity pulses: Most random initialisations lead to successful outcomes after optimisation. 
Thus, we commence our pulse construction by finding long-$T$ pulses. We then use the fine-tuned amplitudes of the segments to initialise the optimisation for the next pulse with shorter $T$. The result is an altered pulse with each segment slightly compressed in duration. Usually, only small adjustments to the segment amplitudes are needed to recover a low-infidelity pulse. In this way, our optimiser iteratively works towards lower values of $T$. At some $T$, the optimiser produces a ``bang-bang'' pulse whose segment amplitudes jump between the minimum and maximum values of the $J$-drives. In these situations, our pulse-contraction methodology breaks down, and we must reinitialise the GRAPE algorithm with a new random pulse.
The data of the $(\theta, T)$-grid presented in Fig.~\ref{fig:thetaTmap} displays the best result out of six different initialisations per probed $(\theta, T)$ coordinate. 
We probe the $\theta$-direction in increments of $\pi/80$ and the $T$-direction in varying (but predefined) increments to increase the resolution close to the MET line. Around the MET line, we use a $T$ increment of 20~ns. At large $T$, the increments approach 70~ns. 

\textit{Data generation for a continuous family of pulses:---}We also investigate whether there exist pulse parameterisations that change continuously with changes in $\theta$. 
For this purpose, we modified our methodology. The data from our analysis is presented in Fig.~\ref{fig:interpolationquality}. First, we explain how we obtain a set of baseline pulses (marked by black dots in  Fig.~\ref{fig:interpolationquality}). Then, we explain how we interpolate between these pulses. 

Figure~\ref{fig:thetaTmap} clarifies that the qubit-excitations' METs are nearly constant over changes in $\theta$. Therefore, we fix $T$ in our $\theta$-dependence analysis. 
For the double-qubit excitation, we choose $T=1020$~ns, well above the MET\@. 
For fixed $\theta$ and $T$, there may exist several pulses that approximate the qubit excitation with low infidelity. To facilitate accurate interpolation, it is useful if the baseline pulses for neighbouring $\theta$ values are similar (\textit{cf.}~Fig.~\ref{fig:continuouspulses}). In terms of Fig.~\ref{fig:interpolationquality}, one desires the pulse represented by one black dot to look similar to the pulses of its neighbouring dots.
To maximise the chances of finding a similar pulse, we use the amplitudes of the fine-tuned pulse for $\theta_i$ to initialise the pulse optimisation of the next strength $\theta_{i+1}$. We proceed iteratively along a list of 100 values of strengths $\theta_i$, $i\in\{1,\ldots,100\}$, evenly spaced across the interval $[0,\frac{\pi}{2}]$. We found it suboptimal to initialise the pulse construction at $\theta_1=0$. Instead, we commence with a $\theta$-value somewhere near the middle of the range and iterate both towards larger and smaller $\theta$-values. 
This method is similar to the one described in the previous section. The only difference is that here we propagate fine-tuned pulses in the $\theta$-direction, rather than in the $T$-direction. 
We build our interpolation on the continuously changing set of baseline pulses. A simple linear interpolation between segment amplitudes for the strength $\theta_i$ and $\theta_{i+1}$ is sufficient to obtain a pulse for an intermediate $\theta\in[\theta_i, \theta_{i+1}]$ to within an infidelity of $\mathcal{I}\approx 10^{-6}$.

\bibliographystyle{naturemag}
\bibliography{references}

\appendix
\renewcommand{\thesubsection}{\Alph{section}.\arabic{subsection}}

\end{document}


\title{Supplementary material: Efficient gate-based qubit excitations in silicon}

\begin{abstract}
In this supplementary material, we express qubit‑excitation circuits using a native gate set comprised of controlled $Z$‑rotations ($\CZ$), an $\Xpih$ gate, and virtual $Z$‑rotations. This native gate set is further augmented by a resonant $\SWAP$ gate. We subsequently discuss the implementation of the $\Xpih$, $\CZ$, and resonant $\SWAP$ gates for Loss-DiVincenzo spin qubits, with the goal of calculating the execution time of each gate. This supplementary material provides supporting information for the execution-time benchmark of gate‑based qubit excitations presented in the main manuscript. The material is based on Ref.~\cite{Rimbach-Russ2023Sep}.
\end{abstract}

\date{\today}
\maketitle

\onecolumngrid

\section{Matrix representations of common qubit gates}

For clarity, we define all gates via their matrix representations in the $Z$-basis. We begin with the standard Pauli matrices
\begin{align}
    I = 
    \begin{pmatrix}
    1 & 0 \\
    0 & 1 \\
    \end{pmatrix},
    \quad
    X = 
    \begin{pmatrix}
    0 & 1 \\
    1 & 0 \\
    \end{pmatrix},
    \quad
    Y = 
    \begin{pmatrix}
    0 & -i \\
    i & 0 \\
    \end{pmatrix},
    \quad
    Z = 
    \begin{pmatrix}
    1 &  0 \\
    0 & -1 \\
    \end{pmatrix}.
\end{align}
General rotations generated by a Pauli matrix $P$ are given by
\begin{align}
    P_\theta = \exp(-i\frac{P}{2}\theta) = I\cos(\frac{\theta}{2}) -iP \sin(\frac{\theta}{2}).
\end{align}
The rotations generated by the standard Pauli matrices are
\begin{align}
    R_x(\theta) = 
    \begin{pmatrix}
    \cos(\theta/2) & -i \sin(\theta/2) \\
    -i \sin(\theta/2) & \cos(\theta/2) \\
    \end{pmatrix},
    \quad
    R_y(\theta) =
    \begin{pmatrix}
    \cos(\theta/2) & -\sin(\theta/2) \\
    \sin(\theta/2) & \cos(\theta/2) \\
    \end{pmatrix},
    \quad
    R_z(\theta) =
    \begin{pmatrix}
    \exp(-i\theta/2) & 0 \\
    0 & \exp(i\theta/2) \\
    \end{pmatrix}.
\end{align}
Another matrix which commonly appears in qubit-excitation circuits is the Hadamard gate,
\begin{align}
    H =
    \frac{1}{\sqrt{2}}
    \begin{pmatrix}
    1 &  1 \\
    1 & -1 \\
    \end{pmatrix}.
\end{align}
A global phase is denoted by
\begin{align}
    Ph(\delta) = I \exp(i\delta).
\end{align}
A special single-qubit gate is given by Ref.~\cite{McKay17},
\begin{align}\label{eq: general}
    U(\theta,\phi,\lambda)=R_z(\phi) R_x(\theta) R_z(\lambda) \mapsto  
    \exp(-i\frac{\lambda+\phi}{2})
    \begin{pmatrix}
    \cos(\theta/2) &
    -i\exp(i\lambda)\sin(\theta/2) \\
    -i\exp(i\phi)\sin(\theta/2) &
    \exp(i[\lambda+\phi])\cos(\theta/2) \\
    \end{pmatrix}.
\end{align}
Importantly, this gate can represent any single‑qubit unitary operation up to an irrelevant global phase. It can further be shown that \cite{McKay17}
\begin{align}
    R_x(\theta) = R_z(-\pi/2) R_x(\pi/2) R_z(\pi-\theta) R_x(\pi/2) R_z(-\pi/2).
\end{align}
Thus, any single‑qubit gate can be expressed using two native $\Xpih$ gates and three virtual $Z$-rotations as
\begin{align}
    U(\theta,\phi,\lambda) =  R_z(\phi-\pi/2) R_x(\pi/2) R_z(\pi-\theta) R_x(\pi/2) R_z(\lambda-\pi/2).
\end{align}
Finally, we give the matrix representations of the $\CZ$ gate, the $\CX$ gate (in the main manuscript referred to as $CNOT$ gate), with the first qubit as the control and the second as the target), the $\SWAP$ gate, and the $\RZZ$ gate:
\begin{align}
    \CZ =
    \begin{pmatrix}
    1 &    &    &    \\
      &  1 &    &    \\
      &    &  1 &    \\
      &    &    & -1
    \end{pmatrix},
    \quad
    \CX =
    \begin{pmatrix}
    1 &    &    &    \\
      &  1 &    &    \\
      &    &    &  1 \\
      &    &  1 & 
    \end{pmatrix},
    \quad
    \SWAP =
    \begin{pmatrix}
    1 &    &    &    \\
      &    &  1 &    \\
      &  1 &    &    \\
      &    &    &  1
    \end{pmatrix},
    \quad
    \RZZ(\theta) =
    \begin{pmatrix}
    e^{-i\theta/2} &    &    &    \\
        & e^{i\theta/2} &    &    \\
        &    & e^{i\theta/2} &    \\
        &    &    & e^{-i\theta/2}
    \end{pmatrix}.
\end{align}
The native gate set in silicon consists of the $\CZ$ gate, the $\Xpih$ gate, and the virtual $Z$ gate. We augment this universal gate set by adding a resonant $\SWAP$ gate \cite{Rimbach-Russ2023Sep}.

\section{Gate-based qubit excitation in silicon}

In this section, we convert the $\CX$-based representation of qubit excitations into a $\CZ$‑based representation. Each single-qubit operation is expressed as a combination of three virtual $Z$ gates and two $\Xpih$ pulses.

\subsection{Single-qubit excitation in silicon}

We start from the standard circuit representation of the single-qubit excitation, as presented in the main body of the manuscript, and replace the $\CX$ gates using the following circuit identity:
\begin{align}
    \begin{quantikz}[align equals at=1.5]
        \qw &
        \ctrl{1} & 
        \qw \\
        \ghost{X} &
        \targ{} & 
        \ghost{Z} \\
    \end{quantikz}
    = 
    \begin{quantikz}[align equals at=1.5]
        \qw &
        \qw &
        \ctrl{1} & 
        \qw &
        \qw \\
        \qw &
        \gate{H} &
        \gate{Z} & 
        \gate{H} & 
        \qw \\
    \end{quantikz}.
\end{align}
Applying this circuit identity yields the single-qubit excitation circuit
\begin{align}
    \begin{quantikz}
        \lstick{$q_k$} & 
        \gate{R_z(\pih)} & 
        \gate{R_x(\pih)} & 
        \ctrl{1} & 
        \qw &
        \gate{R_x(\theta)}  & 
        \qw &
        \ctrl{1} & 
        \gate{R_x(-\pih)} & 
        \gate{R_z(-\pih)} & 
        \qw \\
        \lstick{$q_i$} &
        \gate{R_x(\pih)} & 
        \gate{H} &
        \gate{Z} & 
        \gate{H} &
        \gate{R_z(\theta)} & 
        \gate{H} &
        \gate{Z} & 
        \gate{H} &
        \gate{R_x(-\pih)} & 
        \qw \\
    \end{quantikz}.
\end{align}
Next, we use the following equivalences, which can be derived from the matrix representations:
\begin{align}
    \begin{quantikz}[align equals at=1.0]
        \qw              &
        \gate{R_z(\pih)} & 
        \gate{R_x(\pih)} &
        \qw \\
    \end{quantikz}
    & =
    \begin{quantikz}[align equals at=1.0]
        \qw              &
        \gate{U(\pih, 0, \pih)} &
        \qw \\
    \end{quantikz},
\\
    \begin{quantikz}[align equals at=1.0]
        \qw              &
        \gate{R_x(\pih)} & 
        \gate{H} &
        \qw \\
    \end{quantikz}
    & =
    \begin{quantikz}[align equals at=1.0]
        \qw              &
        \gate{Ph(\pih)} &
        \gate{U(\pih, \pi, \pih)} &
        \qw \\
    \end{quantikz},
\\
    \begin{quantikz}[align equals at=1.0]
        \qw              &
        \gate{R_x(\theta)} & 
        \qw \\
    \end{quantikz}
    & =
    \begin{quantikz}[align equals at=1.0]
        \qw              &
        \gate{U(\theta, 0, 0)} &
        \qw \\
    \end{quantikz},
\\
    \begin{quantikz}[align equals at=1.0]
        \qw                &
        \gate{H}           &
        \gate{R_z(\theta)} &
        \gate{H}           &
        \qw \\
    \end{quantikz}
    & =
    \begin{quantikz}[align equals at=1.0]
        \qw              &
        \gate{U(\theta, 0, 0)} &
        \qw \\
    \end{quantikz},
\\
    \begin{quantikz}[align equals at=1.0]
        \qw                &
        \gate{R_x(-\pih)}  &
        \gate{R_z(-\pih)}  &
        \qw \\
    \end{quantikz}
    & =
    \begin{quantikz}[align equals at=1.0]
        \qw              &
        \gate{U(\pih, \pih, -\pi)} &
        \qw \\
    \end{quantikz},
\\
    \begin{quantikz}[align equals at=1.0]
        \qw                &
        \gate{H}  &
        \gate{R_x(-\pih)}  &
        \qw \\
    \end{quantikz}
    & =
    \begin{quantikz}[align equals at=1.0]
        \qw              &
        \gate{Ph(\pih)}  &
        \gate{U(\pih, \pih, 0)} &
        \qw \\
    \end{quantikz}.
\end{align}
Omitting irrelevant global phases, the single-qubit excitation circuit for silicon devices becomes
\begin{align}
    \begin{quantikz}
        \lstick{$q_k$} & 
        \gate{U (\pih, 0, \pih)} & 
        \ctrl{1} & 
        \gate{U (\theta, 0, 0)}  & 
        \ctrl{1} & 
        \gate{U (\pih,\pih, -\pi)} &
        \qw 
        \\
        \lstick{$q_i$} &
        \gate{U (\pih,\pi,\pih)}&
        \gate{Z} & 
        \gate{U (\theta,0,0)} & 
        \gate{Z} & 
        \gate{U (\pih,\pih, 0)} &
        \qw 
        \\
    \end{quantikz}.
\end{align}

\subsection{Double-qubit excitation in silicon}

Starting from the circuit representation of the double‑qubit excitation given in the main manuscript, we apply the following manipulations. As before, we replace the $\CX$ gates with equivalent circuits based on $\CZ$ and Hadamard ($H$) gates. Wherever possible, consecutive Hadamard gates acting on the same qubit line are eliminated using $H^2=I$. This procedure yields
\begin{center}
    \scalebox{0.56}{
    \begin{quantikz}[column sep=0.2cm, row sep=0.5cm]
        \lstick{$q_l$} & 
        \qw      &
        \ctrl{1} &
        \qw      &
        \qw      &
        \ctrl{2} & 
        \gate{R_y\left(\frac{\theta}{8}\right)} &
        \ctrl{1} & 
        \gate{R_y\left(-\frac{\theta}{8}\right)} & 
        \ctrl{3} &
        \gate{R_y\left(\frac{\theta}{8}\right)} &
        \ctrl{1} &
        \gate{R_y\left(-\frac{\theta}{8}\right)} & 
        \ctrl{2} &
        \gate{R_y\left(\frac{\theta}{8}\right)} & 
        \ctrl{1} & 
        \gate{R_y\left(-\frac{\theta}{8}\right)}& 
        \ctrl{3} & 
        \gate{R_y\left(\frac{\theta}{8}\right)} & 
        \ctrl{1} &
        \gate{R_y\left(-\frac{\theta}{8}\right)} & 
        \qw      & 
        \ctrl{2} & 
        \qw      & 
        \gate{R_z\left(\frac{\pi}{2}\right)}  & 
        \qw      & 
        \qw      & 
        \ctrl{1} & 
        \qw      &
        \qw      \\
        \lstick{$q_k$} &
        \gate{H} &
        \gate{Z} &
        \gate{H} &
        \gate{X} &
        \qw      & 
        \qw      & 
        \gate{Z} & 
        \qw      &
        \qw      &
        \qw      &
        \gate{Z} &
        \qw      &
        \qw      &
        \qw      &
        \gate{Z} &
        \qw      &
        \qw      &
        \qw      &
        \gate{Z} & 
        \qw      &
        \qw      &
        \qw      & 
        \qw      & 
        \qw      &
        \gate{X} &
        \gate{H} &
        \gate{Z} &
        \gate{H} &
        \qw      \\
        \lstick{$q_j$} &
        \qw      &
        \ctrl{1} & 
        \qw      &
        \gate{H} &
        \gate{Z} &
        \gate{H} &
        \qw      &
        \qw      & 
        \qw      & 
        \qw      &
        \qw      &
        \qw      & 
        \gate{Z} &
        \gate{H} & 
        \qw      &
        \qw      &
        \qw      &
        \qw      &
        \qw      &
        \gate{R_z\left(-\frac{\pi}{2}\right)} &
        \gate{H} &
        \gate{Z} &
        \gate{H} &
        \gate{R_z\left(-\frac{\pi}{2}\right)} & \gate{R_y\left(-\frac{\pi}{2}\right)} &
        \qw      & 
        \ctrl{1} &
        \qw      &
        \qw      \\
        \lstick{$q_i$} &
        \gate{H} &
        \gate{Z} &
        \gate{H} &
        \gate{X} &
        \qw      &
        \qw      &
        \qw      & 
        \qw      & 
        \gate{Z} &
        \qw      &
        \qw      &
        \qw      &
        \qw      &
        \qw      &
        \qw      &
        \qw      &
        \gate{Z} &
        \qw      &
        \qw      &
        \qw      &
        \qw      &
        \qw      &
        \qw      &
        \qw      &
        \gate{X} &
        \gate{H} &
        \gate{Z} & 
        \gate{H} &
        \qw      \\
    \end{quantikz}.
    }
\end{center}
To further simplify this circuit, we use the following identities, which can be verified using the matrix representations:
\begin{align}
    \begin{quantikz}[align equals at=1.0]
        \qw              &
        \gate{H}         &
        \qw \\
    \end{quantikz}
    & =
    \begin{quantikz}[align equals at=1.0]
        \qw              &
        \gate{Ph(\pih)} &
        \gate{U(\pih, \pih, \pih)} &
        \qw \\
    \end{quantikz},
\\
    \begin{quantikz}[align equals at=1.0]
        \qw              &
        \gate{H} & 
        \gate{X} &
        \qw \\
    \end{quantikz}
    & =
    \begin{quantikz}[align equals at=1.0]
        \qw              &
        \gate{U(\pih, \pih, -\pih)} &
        \qw \\
    \end{quantikz},
\\
    \begin{quantikz}[align equals at=1.0]
        \qw                &
        \gate{X}  &
        \gate{H}  &
        \qw \\
    \end{quantikz}
    & =
    \begin{quantikz}[align equals at=1.0]
        \qw              &
        \gate{U(\pih, -\pih, \pih)} &
        \qw \\
    \end{quantikz},
\\
    \begin{quantikz}[align equals at=1.0]
        \qw              &
        \gate{R_y(\theta)} & 
        \qw \\
    \end{quantikz}
    & =
    \begin{quantikz}[align equals at=1.0]
        \qw              &
        \gate{U(\theta, \pih, -\pih)} &
        \qw \\
    \end{quantikz},
\\
    \begin{quantikz}[align equals at=1.0]
        \qw                &
        \gate{H}           &
        \gate{R_z(-\pih)} &
        \gate{H}           &
        \qw \\
    \end{quantikz}
    & =
    \begin{quantikz}[align equals at=1.0]
        \qw              &
        \gate{U(-\pih, 0, 0)} &
        \qw \\
    \end{quantikz},
\\
    \begin{quantikz}[align equals at=1.0]
        \qw                &
        \gate{H}           &
        \gate{R_z(-\pih)}  &
        \gate{R_y(-\pih)}  &
        \qw \\
    \end{quantikz}
    & =
    \begin{quantikz}[align equals at=1.0]
        \qw              &
        \gate{Ph(-\pih)} &
        \gate{U(\pih, 0, -\pi)} &
        \qw \\
    \end{quantikz},
\\
    \begin{quantikz}[align equals at=1.0]
        \qw                &
        \gate{U(\theta,\phi,\lambda)}           &
        \gate{R_z(\alpha)}  &
        \qw \\
    \end{quantikz}
    & =
    \begin{quantikz}[align equals at=1.0]
        \qw              &
        \gate{U(\theta, \phi+\alpha, \lambda)} &
        \qw \\
    \end{quantikz}.
\end{align}
Using these identities, omitting irrelevant global phases, and exploiting the fact that $\CZ$ and $R_z$ gates commute, we obtain the following representation of the double‑qubit excitation circuit:
\begin{center}
    \scalebox{0.47}{
    \begin{quantikz}[column sep=0.2cm, row sep=0.5cm]
        \lstick{$q_l$} & 
        \qw      &
        \ctrl{1} &
        \qw      &
        \ctrl{2} & 
        \gate{U(\theight, \pih,-\pih)} &
        \ctrl{1} & 
        \gate{U(-\theight, \pih,-\pih)} & 
        \ctrl{3} &
        \gate{U(\theight, \pih,-\pih)} &
        \ctrl{1} &
        \gate{U(-\theight, \pih,-\pih)} & 
        \ctrl{2} &
        \gate{U(\theight, \pih,-\pih)} & 
        \ctrl{1} & 
        \gate{U(-\theight, \pih,-\pih)}& 
        \ctrl{3} & 
        \gate{U(\theight, \pih,-\pih)} & 
        \ctrl{1} &
        \gate{U(-\theight, \pi,-\pih)} & 
        \ctrl{2} & 
        \qw      & 
        \ctrl{1} & 
        \qw      &
        \qw      \\
        \lstick{$q_k$} &
        \gate{U(\pih,\pih,\pih)} &
        \gate{Z} &
        \gate{U(\pih,\pih,-\pih)} &
        \qw      & 
        \qw      & 
        \gate{Z} & 
        \qw      &
        \qw      &
        \qw      &
        \gate{Z} &
        \qw      &
        \qw      &
        \qw      &
        \gate{Z} &
        \qw      &
        \qw      &
        \qw      &
        \gate{Z} & 
        \qw      &
        \qw      &
        \gate{U(\pih,-\pih,\pih)} &
        \gate{Z} &
        \gate{U(\pih,\pih,\pih)} &
        \qw      \\
        \lstick{$q_j$} &
        \qw      &
        \ctrl{1} & 
        \gate{U(\pih,\pih,\pih)} &
        \gate{Z} &
        \gate{U(\pih,\pih,\pih)} &
        \qw      &
        \qw      & 
        \qw      & 
        \qw      &
        \qw      &
        \qw      & 
        \gate{Z} &
        \qw      &
        \qw      &
        \qw      &
        \qw      &
        \qw      &
        \qw      &
        \gate{U(-\pih,0,0)} &
        \gate{Z} &
        \gate{U(\pih,0,-\pi)} &
        \ctrl{1} &
        \qw      &
        \qw      \\
        \lstick{$q_i$} &
        \gate{U(\pih,\pih,\pih)}  &
        \gate{Z} &
        \gate{U(\pih,\pih,-\pih)} &
        \qw      &
        \qw      &
        \qw      & 
        \qw      & 
        \gate{Z} &
        \qw      &
        \qw      &
        \qw      &
        \qw      &
        \qw      &
        \qw      &
        \qw      &
        \gate{Z} &
        \qw      &
        \qw      &
        \qw      &
        \qw      &
        \gate{U(\pih,-\pih,\pih)} &
        \gate{Z} & 
        \gate{U(\pih,\pih,\pih)} &
        \qw      \\
    \end{quantikz}.
    }
\end{center}

\section{Execution times of gate-based qubit excitations}

From the qubit‑excitation circuits, we find that a single‑qubit excitation requires the execution of 2 layers of $\CZ$ gates and 3 layers of single-qubit gates. Similarly, a double-qubit excitation requires 11 layers of $\CZ$ gates and 12 layers of single-qubit gates. The execution time of a layer of $\CZ$ gates is given by $T_{\CZ}$, the execution time of the $\CZ$ gate. The execution time of each single-qubit layer is given by $2T_{\Xpih}$, corresponding to the execution time of two $\Xpih$ gates (the three $\VZ$ gates are virtual and therefore require no execution time). Thus, the total execution times of a single- and a double-qubit excitation are given by
\begin{align}
    T_{SQ} 
    & = 2 \cdot T_{\CZ} + 3 \cdot 2 \cdot T_{\Xpih}, \\
    T_{DQ}
    & = 11 \cdot T_{\CZ} + 12 \cdot 2 \cdot T_{\Xpih} + 6 \cdot T_{\SWAP}.
\end{align}
We note that the execution time of six $\SWAP$ gates is added to the double-qubit excitation to account for the operation of a linear chain of spin qubits. Specifically, executing a double-qubit excitation requires a $\SWAP$ gate between qubits $k$ and $l$ at the beginning and end of each double-qubit excitation. In addition, a $\SWAP$ gate must be applied before and after each of the two $\CZ$ gates acting between qubits $i$ and $l$ of the double-qubit excitation. 

The remainder of this supplementary material is dedicated to analysing the execution times of the $\Xpih$-gate, the $\CZ$-gate and the $\SWAP$-gate, following the methods outlined in Ref.~\cite{Rimbach-Russ2023Sep}.
The analysis assumes device parameters 
$\Omega_{\max}/h=5$MHz, $J_{\max}/h=10$MHz and detunings in the range $\Delta/h\in[2,300]$MHz.
These values are identical to the device parameters used in the main manuscript to ensure a fair comparison. The analysis yields the execution times summarised in Tab.~\ref{tab:GateExecutionTimes}. The benchmark used in the main manuscript is based on shaped pulses, as these are most likely to be employed in state‑of‑the‑art implementations \cite{Rimbach-Russ2023Sep}.
\begin{table}[h!]
\begin{center}
\begin{tabular}{ |c|c|c||c|c|c||c|c| } 
    \hline
        \textbf{Method} & 
        \textbf{Window} &
        \textbf{Detuning} (MHz) &
        $T_{\Xpih}$ (ns) & 
        $T_{\CZ}$ (ns) & 
        $T_{\SWAP}$ (ns) & 
        $T_{SQ}$ (ns) &
        $T_{DQ}$ (ns) \\ 
    \hline
        Theoretically fastest   & 
        Square &
        --   &
        $50$ & 
        $50$ & 
        $100$ &
        $400$ &
        $2350$ \\ 
    \hline
        \multirow{2}{*}{Synchronisation} & 
        \multirow{2}{*}{Square} &
        $\Delta/h\in[2,300]$ &
        $50$--$484$ & 
        $50$--$433$ &
        $\ge100$ &
        $\ge 400$--$3770$ &
        $\ge 2350$--$16979$ \\ 
    \cline{3-8}
         &
         &
        $\Delta/h=8$ &
        $121$        &
        $108$        &
        $\ge100$     &
        $942$        &
        $\ge 4692$   \\ 
    \hline
        \multirow{4}{*}{Shaped pulses} & 
        \multirow{2}{*}{Kaiser} &
        $\Delta/h=2$ &
        $ \ge 453$   & 
        $ \ge 247$   & 
        $ \ge 100$   &
        $ \ge 3212$  &
        $ \ge 14189$ \\
    \cline{3-8}
         & 
         &
        $\Delta/h=8$ &
        $ \ge 259$ & 
        $ \ge 196$   & 
        $ \ge 100$   &
        $ \ge 1946$  &
        $ \ge 8972$  \\
    \cline{2-8}
             & 
        Hann &
        $\Delta/h=100$ &
        $ \ge 100$     & 
        $ \ge 98$      &
        $ \ge 100$     &
        $ \ge 796$    &
        $ \ge 4078$     \\
    \cline{2-8}
              & 
        Tukey &
        $\Delta/h=300$ &
        $ \ge 56$      & 
        $ \ge 55$      & 
        $ \ge 100$     &
        $ \ge 446 $    &
        $ \ge 2549 $    \\
    \hline
\end{tabular}
\end{center}
\caption{Summary of execution times derived in this supplementary material.
}\label{tab:GateExecutionTimes}
\end{table}

\section{Execution time of the $\Xpih$-gate}

To estimate the execution time $T_{\Xpih}$, we consider the Schr{\"o}dinger equation
\begin{align}
    i\hbar\partial_t \phi(t)= H(t)\phi(t),
\end{align}
for the Heisenberg Hamiltonian of two spins driven by a shared microwave line,
\begin{align}
    H(t)
    =
    -\frac{1}{2}B_1 Z_1-\frac{1}{2}B_2 Z_2+
    \Omega(t) \cos(\omega [t-t_0] - \varphi) \left[X_1 +  X_2\right].
\end{align}
Here, $\Omega(t)$ is the amplitude of the microwave drive. $\Omega(t)\ge0$ is nonzero if and only if $t\in[t_0, t_0+T]$, \textit{i.e.}, for the duration $T$ of the microwave pulse.
Moreover, $\omega$ is the carrier frequency of the microwave pulse, while $\varphi$ denotes its phase.

As we wish to drive the first qubit, we set the drive frequency $\hbar\omega=B_1$ and switch to the rest frame of the two spins, where the qubit is defined.
Using the transformation
\begin{align}
\phi'(t)=R(t)\phi(t),
\end{align}
with
\begin{align}
R(t)=\exp(-i\frac{B_1 Z_1 + B_2 Z_2}{2\hbar}[t-t_0]),
\end{align}
and applying the rotating-wave approximation, we obtain the Schr{\"o}dinger equation
\begin{align}
    i\hbar\partial_t \phi'(t)= H'(t)\phi'(t),
\end{align}
where the Hamiltonian is
\begin{align}
    H'(t)
    =
    \frac{\Omega(t)}{2}
    \left[
    X_1 \cos(\varphi) 
    + Y_1 \sin(\varphi)
    \right]
    +
    \frac{\Omega(t)}{2}
    \left[
    X_2 \cos(\frac{\Delta}{\hbar}[t-t_0] + \varphi) 
    + Y_2 \sin(\frac{\Delta}{\hbar}[t-t_0] + \varphi)\right],
\end{align}
and the detuning energy is $\Delta=B_2-B_1$. 
We see that the microwave drive implements the intended unitary evolution on the first qubit, leading to the gate
\begin{align}
    \label{eq:Xpih}
    G(\theta(T),\varphi) 
    := 
    \exp(-i\frac{\theta}{2}\left[X\cos(\varphi)+Y\sin(\varphi)\right])
    =
    R_z(\varphi) R_x(\theta(T)) R_z(-\varphi)
    ,
\end{align}
with
\begin{align}
    \label{eq:PulseAreaXpih}
    \theta(T) = \int_{t_0}^{t_0+T}\mathrm{d}t\frac{\Omega(t)}{\hbar}.
\end{align}
The gate $G(\theta,\varphi)$ implements the $\Xpih$ gate for pulses satisfying $\theta(T)=\pi/2$. The microwave phase $\varphi$ is used to implement virtual $\VZ$ rotations \cite{McKay17}. Given that $\Omega(t)\le\Omega_{\max}$, this lower bounds the execution time of the $\Xpih$ gate
\begin{align}
    \label{eq:XpihFastest}
    T_{\Xpih}
    \ge
    T_{\Xpih}^{\min}
    :=
    \frac{h}{4\Omega_{\max}}.
\end{align}
Thus, assuming a rectangle pulse and $\Omega_{\max}/h=5$MHz \cite{Philips2022}, the execution time of the $\Xpih$ gate can be as short as $T_{\Xpih}^{\min}=50$ nanoseconds.
In practice, the execution time is often longer. One reason for choosing a slower execution time $T_{\Xpih}$ is the need to reduce crosstalk on other qubits driven by the same microwave signal \cite{Rimbach-Russ2023Sep}. This is discussed in more detail below.

\section{Execution time of the $\Xpih$-gate with crosstalk compensation}

When driving a target qubit via a shared microwave line, other qubits---referred to as \textit{spectator qubits}---experience an unintended drive. To minimise the effect of such unintended drives on the overall fidelity, pulses that mitigate crosstalk must be applied \cite{Rimbach-Russ2023Sep}. This necessarily leads to longer execution times for the $\Xpih$ gate, as discussed in more detail in this section.

\subsection{Time-evolution operator of the spectator qubit}

In the following, we refer to the spectator qubit (\textit{i.e.}, the wavefunction $\phi_2'(t)$ in $\phi'(t) = \phi_1'(t) \otimes \phi_2'(t)$) using the notation
\begin{align}
    \psi_1(t)\equiv\phi_2'(t).
\end{align}
Here, the subscript on $\psi$ indicates that we consider the spectator qubit in the first rotating frame. Dropping subscripts from the spectator-qubit operators then yields the corresponding Schr{\"o}dinger equation
\begin{align}
    \label{eq:SchroedingerSpectatorQubit}
    i\hbar\partial_t \psi_1(t)
    =
    \frac{\Omega(t)}{2}
    \left[
    X \cos(\frac{\Delta}{\hbar}[t-t_0]+\varphi) 
    + Y \sin(\frac{\Delta}{\hbar}[t-t_0]+\varphi)\right] \psi_1(t),
\end{align}
which can be written equivalently as
\begin{align}
    \label{eq:SchroedingerSpectatorQubit}
    i\hbar\partial_t \psi_1(t)
    = \frac{\Omega(t)}{2} R_z(\gamma_t) X R_z(-\gamma_t)\psi_1(t)
    \quad \text{with} \quad
    \gamma_t := \frac{\Delta}{\hbar}[t-t_0]+\varphi.
\end{align}
To solve this equation, we apply a second frame transformation,
\begin{align}
    \psi_2(t) := R_z(-\gamma_t)\psi_1(t).
\end{align}
This transformation yields the Schr{\"o}dinger equation
\begin{align}
    i\hbar\partial_t \psi_2(t)
    & =
    \left[
    -\frac{\Delta}{2} Z +
    \frac{\Omega(t)}{2} X 
    \right] \psi_2(t).
\end{align}
At each time $t$, the Hamiltonian can be expressed in diagonal form, resulting in
\begin{align}
    i\hbar \partial_t \psi_2(t)
    & =
    R_y(\alpha_t) \frac{\Delta'_t}{2} Z R_y(-\alpha_t) \psi_2(t),
\end{align}
where we defined (with $\sigma_\Delta$ denoting the sign of $\Delta$)
\begin{align}
    \alpha_t
    & := -\sigma_{\Delta} \arctan(\frac{\Omega(t)}{|\Delta|}), 
    \label{eq:def_alpha}
    \\
    \Delta_t'&:= -\sigma_{\Delta} \sqrt{\Delta^2 + \Omega^2(t)}.
    \label{eq:def_Delta_Xpih}
\end{align}
Next, we apply a third frame transformation,
\begin{align}
    \psi_3(t) := R_y(-\alpha_t)\psi_2(t),
\end{align}
which results in the Schr{\"o}dinger equation
\begin{align}
    i\hbar \partial_t \psi_3(t)
    & =
    \left[
    \frac{\Delta'_t}{2} Z 
    - \frac{\dot{\alpha}_t\hbar}{2} Y
    \right]
    \psi_3(t).
\end{align}
We then apply a fourth frame transformation,
\begin{align}
    \psi_4(t) = R_z(-\beta_t) \psi_3(t),
\end{align}
with
\begin{align}
    \label{eq:def_beta_Xpih}
    \beta_t 
    :=\int_{t_0}^t\mathrm{d}\tilde{t}
    \frac{\Delta'_{\tilde{t}}}{\hbar}.
\end{align}
This leads to the time evolution
\begin{align}
\label{eq:SingleQubitSchroedingerFourthFrame}
    i\partial_t \psi_4(t)
    & =
    - \frac{\dot{\alpha}_t}{2} 
    R_z(-\beta_t) Y R_z(\beta_t)
    \psi_4(t)
    =
    - \frac{\dot{\alpha}_t}{2}
    \left[
    Y \cos(\beta_t) + X\sin(\beta_t)
    \right]
    \psi_4(t).
\end{align}
Combining all frame transformations, we obtain
\begin{align}
    \psi_4(t) = R_z(-\beta_t) R_y(-\alpha_t) R_z(-\gamma_t)\psi_1(t).
\end{align}
Moreover, the time evolution in the fourth frame can be expressed in terms of a time‑evolution operator $U_4(t,t_0)$
\begin{align}
    \psi_4(t) = U_4(t,t_0) \psi_4(t_0).
\end{align}
Combining both results, the time-evolution operator that evolves the electron spin in the qubit frame,
$\psi_1(t)=U_1(t,t_0)\psi_1(t_0)$ is given by
\begin{align}
    \label{eq:SpectatorUnitary}
    U_1(t,t_0) = 
    R_z(\gamma_t) R_y(\alpha_t) R_z(\beta_t)
    U_4(t,t_0) R_z(-\beta_{t_0}) R_y(-\alpha_{t_0}) R_z(-\gamma_{t_0}).
\end{align}

\subsection{Fidelity of the $\Xpih$-gate with crosstalk}

In what follows, we apply the strategy of Ref.~\cite{Rimbach-Russ2023Sep} to ensure that the unintended unitary evolution of the spectator qubit does not degrade the overall fidelity of the system. The general strategy is to engineer a unitary evolution of the spectator qubit that is either (i) equivalent to the identity or (ii) a known $Z$-rotation, as the latter can be readily undone using virtual gates. This equivalence must hold up to an irrelevant global phase $\zeta$.

More specifically, we aim to implement the target unitary
\begin{align}
    U_i 
    := 
    G(\theta,\varphi) \otimes R_z(\chi).
\end{align}
By contrast, the actual unitary implemented over the gate‑execution time $T_{\Xpih}$  is given by
\begin{align}
    U(t_0+T_{\Xpih}, t_0)
    := 
    G(\theta,\varphi) \otimes U_1(T_{\Xpih}).
\end{align}
We therefore seek to design a pulse sequence, that minimises $\epsilon$ in the undesired evolution, which, as we will see, takes the form
\begin{align}
    U_1(t_0+T_{\Xpih},t_0)=
    R_z(\eta) 
    \exp(-i\frac{1}{2} H_{\epsilon}) 
    \exp(i\zeta)
    R_z(\xi),
\end{align}
Here, we assume a small perturbation
\begin{align}
    H_{\epsilon} 
    :=
    \begin{pmatrix}
          & \epsilon \\
    \epsilon^* &  
    \end{pmatrix}
    =
    |\epsilon|
    \begin{pmatrix}
          & \exp(i\arg(\epsilon)) \\
    \exp(-i\arg(\epsilon)) &  
    \end{pmatrix}
    =:
    |\epsilon| P_{\epsilon}.
\end{align}
We further note that the angles $\eta$ and $\xi$ determine the phase $\chi:=\eta+\xi$. Using the average gate fidelity formula \cite{NIELSEN2002249} for a Hilbert space comprising the target and spectator qubits (with Hilbert-space dimension $d=4$), we obtain
\begin{align}
    F_{\Xpih}
    &:= 
    \frac{|\mathrm{Tr}[U_i^{\dagger} U(t_0+T_{\Xpih},t_0)]|^2+d}{d(d+1)}
    = 
    \frac{|\mathrm{Tr}[1\otimes(1\cos(|\epsilon|/2)
    -i P_{\epsilon}\sin(|\epsilon|/2))]|^2+d}{d(d+1)}.
\end{align}
Evaluating the trace yields
\begin{align}
    F_{\Xpih}
    =
    \frac{|4\cos(|\epsilon|/2)|^2+d}{d(d+1)}
    =
    \frac{4 \cos^2(|\epsilon|/2)+1}{(d+1)}
    =
    \frac{5 - 4\sin^2(|\epsilon|/2)}{5}.
\end{align}
Conversely, the infidelity is given by
\begin{align}
    \label{eq:InfidelityXpih}
    I_{\Xpih}
    :=
    1-F_{\Xpih} 
    =
    \frac{4 \sin^2(|\epsilon|/2)}{5}
    \approx
    \frac{|\epsilon|^2}{5}.
\end{align}

\subsection{Execution times for the $\Xpih$ gate using rectangular pulses: The synchronisation method}

\begin{figure}
    \centering
    \includegraphics[width=0.5\textwidth]{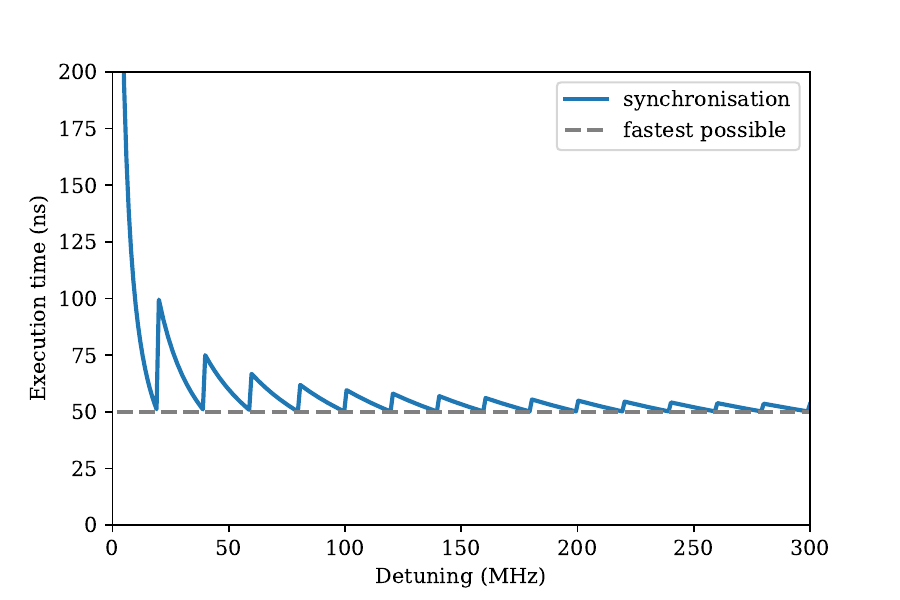}
    \caption{
    Execution time of the $\Xpih$ gate [Eq.~\eqref{eq:SingleQubitGateSynchronisation}, blue line] as a function of detuning $\Delta/h$, assuming $\Omega_{\max}/h = 5$ and the use of the synchronisation method to compensate for crosstalk on a neighbouring spectator qubit. The dashed line indicates the lower bound of Eq.~\eqref{eq:XpihFastest}.
    }\label{fig:OneQubitSynch}
\end{figure}

The first case we consider is driving with a constant rectangular pulse, implying $\Omega(t) = \Omega = \mathrm{const.}$ for the duration $T_{\Xpih}$. In this case, $\dot{\alpha}_{t}=0$, which, via Eq.~\eqref{eq:SingleQubitSchroedingerFourthFrame} implies that $U_4(t,t_0)=I$ for all $t\in[t_0, t_0+T_{\Xpih}]$.
The outer two factors of $U_1(t_0+T_{\Xpih},t_0)$ in Eq.~\eqref{eq:SpectatorUnitary}, $R_z(\gamma_{t_0+T_{\Xpih}})$ and $R_z(-\gamma_{t_0})$, can be absorbed via a $Z$ rotation, which can in turn be implemented using a virtual $\VZ$ gate.
The remaining non-trivial contribution is the central part of $U_1(t_0+T_{\Xpih}, t_0)$, given by
\begin{align}
    \label{eq:XpihCentralSync}
    R_y(\alpha_{t_0+T_{\Xpih}}) R_z(\beta_{t_0+T_{\Xpih}}-\beta_{t_0}) R_y(-\alpha_{t_0}).
\end{align}
Using Eq.~\eqref{eq:def_alpha}, we note that
\begin{align}
    \alpha_{t_0} = \alpha_{t_0+T_{\Xpih}} = -\sigma_{\Delta} \arctan(\frac{\Omega}{|\Delta|})\neq 0.
\end{align}
As a consequence, this central part corresponds to a rotation about an axis tilted away from the standard $Z$-axis. To ensure that this contribution is equivalent to the identity operator (up to a global phase), the gate-execution time $T_{\Xpih}$ must be chosen such that
\begin{align}
    \label{eq:SyncConditionXpih}
    |\beta_{t_0+T_{\Xpih}}-\beta_{t_0}|=2 \pi m,
\end{align}
with integer $m$. When this condition is satisfied, the central part, Eq.~\eqref{eq:XpihCentralSync}, is equivalent to the identity operator up to a global phase.
Thus, imposing the condition in Eq.~\eqref{eq:SyncConditionXpih} yields the gate-execution times [using Eq.~\eqref{eq:def_beta_Xpih} and Eq.~\eqref{eq:def_Delta_Xpih}] 
\begin{align}
    T_{\Xpih}(m) = \frac{hm}{\sqrt{\Delta^2+\Omega^2}}.  
\end{align}
At the same time, the gate‑execution time must be chosen to ensure that the pulse area, Eq.~\eqref{eq:PulseAreaXpih}, is $\pi/2$, leading to the second condition
\begin{align}
    T_{\Xpih} = \frac{h}{4\Omega}.  
\end{align}
To satisfy both conditions simultaneously for a given detuning $\Delta$, the drive amplitude must be chosen as
\begin{align}
    \Omega(m) = \frac{|\Delta|}{\sqrt{16m^2-1}}.
\end{align}
This results in permissible gate-execution times
\begin{align}
    T_{\Xpih}(m) = \frac{h}{4|\Delta|}\sqrt{16m^2-1}.  
\end{align}
In addition, the constraint $\Omega\le\Omega_{\max}$ must be satisfied. This restricts the allowed values of $m$ to the range
\begin{align}
    m \ge \frac{\sqrt{\Delta^2+\Omega_{\max}^2}}{4\Omega_{\max}}.
\end{align}
The minimal value in this range yields the largest allowable $\Omega$ and thus the shortest gate-execution time. This minimal value is given by
\begin{align}
    m_{\min} := \bigg\lceil\frac{\sqrt{\Delta^2+\Omega_{\max}^2}}{4\Omega_{\max}}\bigg\rceil.
\end{align}
Substituting this value back into the gate‑time expression yields the fastest possible gate‑execution time achievable using synchronisation,
\begin{align}
    \label{eq:SingleQubitGateSynchronisation}
    T_{\Xpih}^{\min,\mathrm{syn}} := \frac{h}{4|\Delta|}\sqrt{16 \bigg\lceil\frac{\sqrt{\Delta^2+\Omega_{\max}^2}}{4\Omega_{\max}}\bigg\rceil^2 - 1} 
    \ge \frac{h}{4\Omega_{\max}}=:T_{\Xpih}^{\min}. 
\end{align}
In this work, we consider $\Omega_{\max}/h=5$ MHz and $\Delta/h\in[2,300]$ MHz. Depending on the specific detuning value, this leads to gate-execution times $T_{\Xpih}$ in the range $50$--$484$ nanoseconds, as shown in Fig.~\ref{fig:OneQubitSynch}. Gate-execution times can become particularly long for small detuning values $\Delta$, which are common in devices without micromagnets. In particular, for $\Delta/h=8$ MHz, as used in the main manuscript, the execution time is approximately $121$ nanoseconds.

\subsection{Execution times for the $\Xpih$ gate using shaped pulses}

Achieving high fidelities with rectangular pulses is challenging  \cite{Rimbach-Russ2023Sep}, as (i) even small timing mismatches can lead to errors, and (ii) filtering in the control lines can distort the pulses and introduce infidelities. Crosstalk compensation by synchronisation is also difficult when a common microwave signal drives more than one spectator qubit with distinct detuning values $\Delta_1$ and $\Delta_2$ simultaneously. For these reasons, it is preferable to implement high-fidelity single‑qubit gates using shaped pulses $\Omega(t)$.

\subsubsection{Preliminaries}

When implementing a shaped pulse of duration $T_{\Xpih}$ for the time $t\in[t_0,t_0+T_{\Xpih}]$, we choose pulses satisfying 
\begin{align}
 \Omega(t_0) = \Omega(t_0+T_{\Xpih})=0.
\end{align}
This choice implies, using  Eq.~\eqref{eq:def_alpha}, that $\alpha_{t_0}=\alpha_{t_0+T_{\Xpih}}=0$.
As a result, the unintended unitary evolution induced on the spectator qubit, Eq.~\eqref{eq:SpectatorUnitary}, simplifies to
\begin{align}
    U_1(t_0+T_{\Xpih},t_0) 
    &= 
    R_z(\gamma_{t_0+T_{\Xpih}}) R_y(\alpha_{t_0+T_{\Xpih}}) 
    R_z(\beta_{t_0+T_{\Xpih}})
    U_4({t_0+T_{\Xpih},t_0}) 
    R_z(-\beta_{t_0})
    R_y(-\alpha_{t_0})
    R_z(-\gamma_{t_0}), \\
    &= 
    R_z(\gamma_{t_0+T_{\Xpih}} + \beta_{t_0+T_{\Xpih}})
    U_4(t_0+T_{\Xpih},t_0)
    R_z(-\beta_{t_0}-\gamma_{t_0}).
\end{align}
The $R_z$ operators correspond to known $Z$-rotations and can therefore be absorbed into virtual $\VZ$ gates. Consequently, the fidelity in the presence of crosstalk is fully determined by the time‑evolution operator $U_4(t_0+T_{\Xpih}, t_0)$.

\subsubsection{Dilated time}

To calculate $U_4(t_0+T_{\Xpih},t_0)$, we introduce a dilated time variable
\begin{align}
    \label{eq:def_dilated_time_Xpih}
    s(t) := \frac{-\sigma_{\Delta}\beta_t}{\nu T_{\Xpih}}
    \quad \text{such that:}
    \quad s(t_0)=0
    \quad \text{and}
    \quad s(t_0+T_{\Xpih})=1.
\end{align}
Using this definition, we transform the fourth frame of the spectator qubit, Eq.~\eqref{eq:SingleQubitSchroedingerFourthFrame}, into dilated time by defining
\begin{align}
    \tilde{\psi}_4(s)
    &:=\psi_4(t(s)),\\
    \tilde{U}_4(s)
    &:= U_4(t(s), t_0),\\
    \tilde{\alpha}_s 
    &:= \alpha_{t(s)}=-\sigma_{\Delta}\arctan(\frac{\Omega(t(s))}{|\Delta|}).
\end{align}
This results in the Schr{\"o}dinger equation
\begin{align}
    i\partial_s \tilde{\psi}_4(s)
    & =
    \frac{\dot{\tilde{\alpha}}_s}{2}
    \begin{pmatrix}
          & i\exp(-i\sigma_{\Delta}\nu T_{\Xpih} s) \\
    -i\exp(i\sigma_{\Delta}\nu T_{\Xpih} s) &  
    \end{pmatrix}
    \tilde{\psi}_4(s).
\end{align}
Here, $\dot{\tilde{\alpha}}(s)$ denotes the derivative of $\tilde{\alpha}(s)$ with respect to the dilated time $s$.

\subsubsection{Solving the time-dilated unitary}

Assuming a smooth pulse envelope, the derivative $\dot{\tilde{\alpha}}_s$ is small.
We therefore approximate the time-evolution operator using the first‑order Magnus expansion, yielding
\begin{align}
    U_4(t_0+T_{\Xpih}, t_0) \equiv \tilde{U}_4(1) \approx \exp(-i\bar{H}_1),
\end{align}
with
\begin{align}
    \bar{H}_1
    = \int_{0}^1\mathrm{d}s     \frac{\dot{\tilde{\alpha}}_s}{2}
    \begin{pmatrix}
          & i\exp(-i\sigma_{\Delta}\nu T_{\Xpih} s) \\
    -i\exp(i\sigma_{\Delta}\nu T_{\Xpih} s) &  
    \end{pmatrix}
    =: 
    \frac{1}{2}
    \begin{pmatrix}
          & \epsilon \\
    \epsilon^* &  
    \end{pmatrix}.
\end{align}
Here, we define
\begin{align}
    \label{eq:def_epsilon}
    \epsilon 
    &:= i\int_{0}^1\mathrm{d}s     \dot{\tilde{\alpha}}_s\exp(-i\sigma_{\Delta}\nu T_{\Xpih} s).
\end{align}
We note that this expression for $\epsilon$ is consistent with the infidelity formula in Eq.~\eqref{eq:InfidelityXpih}. Our objective is therefore to identify shaped pulses that yield small values of $|\epsilon|^2$.

\subsubsection{Window functions}

To achieve small values of $|\epsilon|^2$, we follow the pulse‑shaping methods described in Ref.~\cite{Rimbach-Russ2023Sep}. Specifically, we choose a function $\tilde{\alpha}_s$ such that $|\epsilon|^2$ is small. To this end, we introduce a window function $W(s)$, which is a real and positive function over the interval $s\in[0,1]$, with the following properties:
\begin{align}
    W(s)\ge0, 
    \quad W(0)=W(1)=0, 
    \quad \int_0^1 \mathrm{d}s W(s) = 1,
    \quad
    W_{\max}:=\max_{s\in[0,1]} W(s).
\end{align}
Using this window function, we define a pulse in dilated time as
\begin{align}
    \label{eq:def_dilated_pulse}
    \tilde{\Omega}(s) := \Omega(t(s)) = |\Delta|\tan\left[AW(s)+\arctan(\frac{\Omega(t_0)}{|\Delta|})\right]\ge0.
\end{align}
Here, $A$ is a real, positive prefactor that scales the pulse amplitude. 
To set the value of $A$, we note that $\tilde{\Omega}(s)$ takes its maximum value $\Omega_{\max}$, when the window function $W(s)$ attains its maximum value $W_{\max}$.
Therefore,
\begin{align}
    \label{eq:def_A}
    A = 
    \frac{
    \arctan(\frac{\Omega_{\max}}{|\Delta|}) 
    - \arctan(\frac{\Omega(t_0)}{|\Delta|})
    }{W_{\max}}.
\end{align}
To gain some intuition, consider a weak pulse where $\Omega(t_0)\le\Omega_{\max}\ll|\Delta|$. This is indicative of devices with a micromagnet.
In that limit, an expansion of tangent and arc-tangent terms to linear order yields a pulse envelope that is proportional to the window function
\begin{align}
    \tilde{\Omega}(s) \approx  \Omega_{\max} \frac{W(s)}{W_{\max}} +\Omega(t_0)\left(1-\frac{W(s)}{W_{\max}}\right).
\end{align}
For stronger pulses where $|\Delta| \lesssim \Omega_{\max}$ on the other hand, the tangent will distort the shape of the window function.

The motivation for choosing the pulse shape in Eq.~\eqref{eq:def_dilated_pulse} is as follows.
Substituting the pulse shape into Eq.~\eqref{eq:def_alpha} yields the dimensionless quantity
\begin{align}
    \tilde{\alpha}_s 
    = 
    -\sigma_{\Delta} \arctan(\frac{\Omega(t(s))}{|\Delta|})
    =
    -\sigma_{\Delta} 
    \left[
    A W(s) + \arctan(\frac{\Omega(t_0)}{|\Delta|})
    \right].
\end{align}
Using this expression, we calculate $\epsilon$ from Eq.~\eqref{eq:def_epsilon} by integration by parts, obtaining
\begin{align}
    -i\epsilon
    & = 
    \int_{0}^1\mathrm{d}s \dot{\tilde{\alpha}}_s \exp(-i\sigma_{\Delta}\nu T_{\Xpih} s) \\
    &=
    i\sigma_{\Delta}\nu T_{\Xpih} \int_{0}^1\mathrm{d}s  
    \tilde{\alpha}_s \exp(-i\sigma_{\Delta}\nu T_{\Xpih} s)
    + \left.\tilde{\alpha}_s \exp(-i\sigma_{\Delta}\nu T_{\Xpih} s)\right|_{0}^{1}\\
    &=
    i\sigma_{\Delta}\nu T_{\Xpih} \int_{0}^1\mathrm{d}s  
    [-\sigma_{\Delta} A W(s) + \tilde{\alpha}_0] \exp(-i\sigma_{\Delta} \nu T_{\Xpih} s)
    + \left. [-\sigma_{\Delta} A W(s) + \tilde{\alpha}_0] \exp(-i\sigma_{\Delta} \nu T_{\Xpih} s)\right|_{0}^{1} \\
    &= 
    -i\nu T_{\Xpih} A 
    \int_{0}^1\mathrm{d}s  
    W(s) \exp(-i\sigma_{\Delta}\nu T_{\Xpih} s).
\end{align}
Defining the squared norm of the Fourier transform as the signal
\begin{align}
    \label{eq:Signal}
    S[W(\nu T_{\Xpih})] := \left| \int_{0}^1\mathrm{d}s  
    W(s) \exp(i\nu T_{\Xpih} s)\right|^2,
\end{align}
we obtain
\begin{align}
    \label{eq:EpsilonAndSignal}
    |\epsilon|^2 = |\nu T_{\Xpih} A|^2 S[W(\nu T_{\Xpih})].
\end{align}
Thus, to achieve a small infidelity in Eq.~\eqref{eq:InfidelityXpih}, the window function must be chosen such that the signal is small. The design of window functions with small spectral leakage is a well‑established problem in signal processing \cite{Rimbach-Russ2023Sep}. Several examples are discussed below, where we also show how the pulse parameter $\nu$ and the execution time $T_{\Xpih}$ can be calculated.

\subsubsection{Pulse parameter equations}

Next, we require two equations to fix the pulse parameters $A$ and $\nu$. One equation originates from the transformation between dilated and real time, Eq.~\eqref{eq:def_dilated_time_Xpih}. The other equation arises from the pulse‑area requirement of the $\Xpih$ gate, Eq.~\eqref{eq:PulseAreaXpih}.

\textit{Dilated vs real time:---}To make the relation between dilated and real time explicit, we use Eq.~\eqref{eq:def_dilated_time_Xpih} to show that
\begin{align}
    \label{eq:DilatedTimeDifferntial}
    \frac{\mathrm{d}s}{\mathrm{d}t} 
    = 
    \frac{-\sigma_{\Delta}}{\nu T_{\Xpih}} \frac{\mathrm{d}\beta_t}{\mathrm{d}t}
    = 
    \frac{1}{\nu T_{\Xpih} \hbar} \sqrt{\Delta^2+\Omega(t)^2}
    =
    \frac{|\Delta|}{\nu T_{\Xpih} \hbar} \cdot
    \frac{1}{\cos\left[A W(s(t)) + \arctan(\frac{\Omega(t_0)}{|\Delta|}) \right]}.
\end{align}
In the last step, we used the definition of the pulse envelope in dilated time, Eq.~\eqref{eq:def_dilated_pulse}. Integrating this equation yields
\begin{align}
    \frac{t-t_0}{T_{\Xpih}} = \frac{\nu \hbar}{|\Delta|} \int_{0}^{s} \mathrm{d}\tilde{s} \cos\left[A W(\tilde{s}) + \arctan(\frac{\Omega(t_0)}{|\Delta|}) \right]
\end{align}
In particular, when the upper integration limit is $s=1$, we have $t-t_0 = T_{\Xpih} $, and the above equation fixes $\nu$ as
\begin{align}
    \label{eq:TimeDilationCondition}
    \frac{1}{\nu} = \frac{\hbar}{|\Delta|} \int_{0}^{1} \mathrm{d}s \cos\left[A W(s) + \arctan(\frac{\Omega(t_0)}{|\Delta|}) \right].
\end{align}

\textit{Pulse-area equation:---}Moreover, to ensure that the target qubit undergoes an $\Xpih$ gate, the pulse area, Eq.~\eqref{eq:PulseAreaXpih}, must be $\pih$, such that
\begin{align}
    \label{eq:PulseAreaCondition}
    \frac{\pi}{2} 
    = 
    \int_{t_0}^{t_0+T_{\Xpih}}\frac{\Omega(t)}{\hbar}\mathrm{d}t 
    = 
    \int_{0}^{1}\frac{\tilde{\Omega}(s)}{\hbar}
    \left.
    \frac{\mathrm{d}t}{\mathrm{d}s}
    \right|_{t=t(s)}
    \mathrm{d}s
    =
    \nu T_{\Xpih} \int_{0}^{1} \mathrm{d}s
    \sin\left[A W(s)+\arctan(\frac{\Omega(t_0)}{|\Delta|})\right].
\end{align}
To obtain the final equality, we used Eq.~\eqref{eq:DilatedTimeDifferntial} together with the definition of the dilated pulse in Eq.~\eqref{eq:def_dilated_pulse}.
Finally, substituting Eq.~\eqref{eq:TimeDilationCondition} into Eq.~\eqref{eq:PulseAreaCondition} determines the gate-execution time 
\begin{align}
    \label{eq:GateExecutionTimeXpih}
    T_{\Xpih} 
    =
    \frac{h}{4|\Delta|}
    \frac{
    \int_{0}^{1} \mathrm{d}s
    \cos[A W(s)+\arctan(\frac{\Omega(t_0)}{|\Delta|})]
    }{
    \int_{0}^{1} \mathrm{d}s
    \sin[A W(s)+\arctan(\frac{\Omega(t_0)}{|\Delta|})]
    }.
\end{align}
For arbitrary window functions $W(s)$, the conditions arising from dilated time, Eq.~\eqref{eq:TimeDilationCondition}, and the pulse area, Eq.~\eqref{eq:PulseAreaCondition}, can only be solved numerically.
Thus, the next paragraphs, will derive analytical lower bounds on the gate-execution time.

\subsubsection{Lower bound on gate-execution times}

\textit{Preliminary inequalities:---}
To derive lower bounds on the gate-execution time, we first establish some simple inequalities.
First, we define the argument of the trigonometric functions as
\begin{align}
    \label{def:Lambda}
    \Lambda(s)
    := 
    A W(s)+\arctan(\frac{\Omega(t_0)}{|\Delta|})
    \stackrel{Eq.~\eqref{eq:def_A}}{=}
    \left[
    \arctan(\frac{\Omega_{\max}}{|\Delta|})
    -
    \arctan(\frac{\Omega(t_0)}{|\Delta|})
    \right]
    \frac{W(s)}{W_{\max}}
    +
    \arctan(\frac{\Omega(t_0)}{|\Delta|}).
\end{align}
Since $\Omega_{\max}\ge\Omega(t_0)$, $\Lambda(s)$ takes its maximum value when $W(s)=W_{\max}$. Therefore, $\Lambda(s)$ takes values in the interval $[0,\pi/2]$.
For such functions one easily shows that $\sin(\Lambda(s))\le\Lambda(s)$. And thus,
\begin{align}
    \label{eq:int_sin_term_lower}
    \frac{1}{\int_0^1\mathrm{d}s \sin(\Lambda(s))} 
    \ge 
    \frac{1}{\int_0^1\mathrm{d}s \Lambda(s)}
    =
    \frac{W_{\max}}{
    \arctan(\frac{\Omega_{\max}}{|\Delta|})
    +
    (W_{\max}-1)\arctan(\frac{\Omega(t_0)}{|\Delta|})
    }
    \stackrel{\Omega(t_0)\ll\Omega_{\max},|\Delta|}{\approx}
    \frac{W_{\max}}{
    \arctan(\frac{\Omega_{\max}}{|\Delta|})}
    .
\end{align}
Conversely, for $\Lambda\in[0,\pi/2]$ one shows that $2 \Lambda(s)/\pi \le \sin(\Lambda(s))$, and thus 
\begin{align}
    \label{eq:int_sin_term_upper}
    \frac{1}{\int_0^1\mathrm{d}s \sin(\Lambda(s))} 
    \le 
    \frac{\pi}{2}
    \frac{1}{\int_0^1\mathrm{d}s \Lambda(s)}
    =
    \frac{\pi}{2}
    \frac{1}{
    A + \arctan(\frac{\Omega(t_0)}{|\Delta|})
    }
    \le
    \frac{\pi}{2A}.
\end{align}
And finally, since $\cos(\Lambda(s))$ takes its minimum value, when $W(s)=W_{\max}$ one shows that
\begin{align}
    \label{eq:int_cos_term_lower}
    \int_0^1\mathrm{d}s \cos(\Lambda(s))
    \ge 
    \cos\left[\arctan(\frac{\Omega_{\max}}{|\Delta|})\right]
    =
    \frac{|\Delta|}{\sqrt{|\Delta|^2+\Omega_{\max}^2}}
    =
    \frac{|\Delta|}{\Omega_{\max}}
    \frac{\Omega_{\max}}{\sqrt{|\Delta|^2+\Omega_{\max}^2}}
    =
    \frac{|\Delta|}{\Omega_{\max}}
    \sin\left[\arctan(\frac{\Omega_{\max}}{|\Delta|})\right]
    .
\end{align}

\textit{Lower bound on gate-execution time from pulse area:---}
Substituting, Eq.~\eqref{eq:int_cos_term_lower} and Eq.~\eqref{eq:int_sin_term_lower} into Eq.~\eqref{eq:GateExecutionTimeXpih} gives a lower bound of the execution time which originates from the pulse area condition
\begin{align}
    \label{eq:GateTimeAreaXpih}
    T_{\Xpih}
    \gtrsim 
    \frac{h W_{\max}}{4\Omega_{\max}}
    \frac{
    \sin\left[\arctan(\frac{\Omega_{\max}}{|\Delta|})\right]
    }{
    \arctan(\frac{\Omega_{\max}}{|\Delta|})
    }
    = 
    \frac{h W_{\max}}{4\Omega_{\max}}
    \mathrm{sinc}\left(\arctan(\frac{\Omega_{\max}}{|\Delta|})\right)
    =:
    T_{\Xpih}^{\min, \mathrm{area}}.
\end{align}
For large detunings $|\Delta| \gg \Omega_{\max}$ (indicative of devices with a micromagnet), both the sine and arctangent functions are in their linear regime. In this limit, the lower bound simplifies to
\begin{align}
    \label{eq:GateTimeAreaScaledXpih}
    T_{\Xpih}^{\min, \mathrm{area}} 
    \approx 
    \frac{h}{4\Omega_{\max}} W_{\max}
    =:
    T_{\Xpih}^{\min} W_{\max}.
\end{align}
This result is intuitive. For rectangular pulses, $W_{\max}=1$, and we recover the fastest possible gate time given in Eq.~\eqref{eq:XpihFastest}. For normalised shaped pulses, however, $W_{\max}>1$. Since the same pulse area must be achieved with a shaped pulse, a proportionally longer gate-execution time is required.

\textit{Lower bound of the gate time from fidelity:---}
In the previous paragraph, we derived a lower bound on the gate‑execution time $T_{\Xpih}$ from the requirement to achieve a pulse area of $\pi/2$. In addition, we must ensure that the overall fidelity, Eq.~\eqref{eq:InfidelityXpih}, including the spectator qubit, remains high. To this end, we introduce a threshold value $x_{\Xpih}(I_t)$ such that
\begin{align}
    \label{eq:ThresholdXpih}
    \forall \nu T_{\Xpih} \ge x_{\Xpih}(I_t): 
    \quad
    S[W(\nu T_{\Xpih})]\le \frac{80}{\pi^4}I_t.
\end{align}
Choosing this threshold ensures that, for all $\nu T_{\Xpih} \ge x_{\Xpih}(I_t)$, the infidelity is bounded as
\begin{align}
    \label{eq:InfidelityXpihSignal}
    I_{\Xpih}
    \approx
    |\nu T_{\Xpih} A|^2
    \frac{S[W(\nu T_{\Xpih})]}{5}
    =
    \left|
    \frac{A \pi}{2\int_{0}^{1}\sin(\Lambda(s))\mathrm{d}s}
    \right|^2
    \frac{S[W(\nu T_{\Xpih})]}{5}
    \le
    \left|\frac{\pi^2}{4}\right|^2
    \frac{S[W(\nu T_{\Xpih})]}{5}
    \le
    I_t.
\end{align}
To derive the first relation, we substitute Eq.~\eqref{eq:EpsilonAndSignal} into Eq.~\eqref{eq:InfidelityXpih}.
For the second equality, we use the pulse-area condition, Eq.~\eqref{eq:PulseAreaCondition}, and make use of the definition of $\Lambda(s)$, Eq.~\eqref{def:Lambda}.
Subsequently, we use the upper bound in Eq.~\eqref{eq:int_sin_term_upper}, followed by the upper bound on the right hand side of  Eq.~\eqref{eq:ThresholdXpih}.

Finally, the condition $x_{\Xpih}(I_t)\le\nu T_{\Xpih}$ in Eq.~\eqref{eq:ThresholdXpih}, imposes a constraint on the execution time $T_{\Xpih}$, given by
\begin{align}
    \label{eq:GateTimeFidelityXpih}
    T_{\Xpih}
    \ge 
    \frac{x_{\Xpih}(I_t)}{\nu}
    =
    \frac{x_{\Xpih}(I_t) h}{2\pi |\Delta|}\int_{0}^{1}\cos(\Lambda(s))\mathrm{d}s
    \ge
    \frac{x_{\Xpih}(I_t) h}{2\pi \sqrt{\Delta^2+\Omega_{\max}^2}}
    =:
    T_{\Xpih}^{\min, \mathrm{fid}}
    .
\end{align}
Here, we use Eq.~\eqref{eq:TimeDilationCondition} to establish the first equal sign. The second inequality, follows from Eq.~\eqref{eq:int_cos_term_lower}.

\subsubsection{Examples}

\begin{figure}
    \centering
    \includegraphics[width=0.9\textwidth]{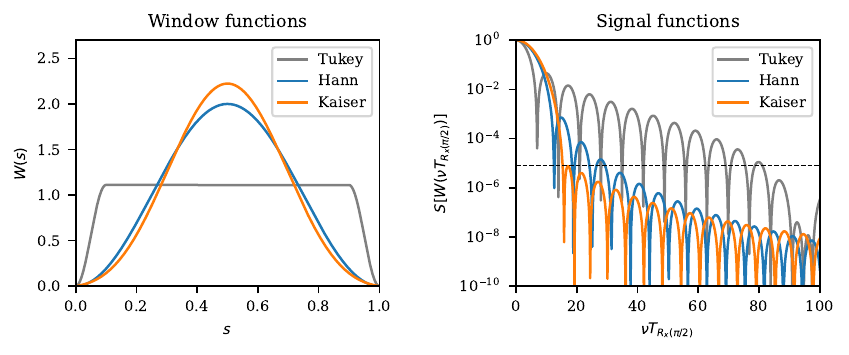}
    \caption{
    (a) Window functions: Hann (blue), Kaiser (orange), and Tukey (grey), with parameters listed in Tab.~\ref{tab:WindowParamsXpih}.
    (b) Signal $S[W(\nu T_{\Xpih})]$ for the different window functions. The dashed line indicates the threshold value $80 I_t/\pi^4$ for a target infidelity $I_t=10^{-5}$.
    }\label{fig:WindowFunctions}
\end{figure}

We now consider three typical window functions, following Ref.~\cite{Rimbach-Russ2023Sep}:\ the Hann window, the Tukey window, and the Kaiser window. Recall that a window function is non‑zero only on the interval $s\in[0,1]$.
First, the Hann window is defined as
\begin{align}
    \label{eq:def_Hann}
    W(s) = 1 - \cos(2\pi s).
\end{align}
The Tukey window is defined as
\begin{align}
    \label{eq:def_Tukey}
    W_{\lambda}(s) = 
    \begin{cases}
        \frac{1}{2-\lambda} 
        \left(1 - \cos(\frac{2\pi s}{\lambda} )\right)
        & 
        \text{if } s < \lambda/2,
        \\
        \frac{2}{2-\lambda} 
        & 
        \text{if } \lambda/2 \le s \le 1 - \lambda/2,
        \\
        \frac{1}{2-\lambda} 
        \left(1 - \cos(\frac{2\pi (1-s)}{\lambda} )\right)
        & 
        \text{if } 1-\lambda/2 \le s.
\end{cases}
\end{align}
The Tukey window interpolates between a rectangular pulse ($\lambda=0$) and a Hann window ($\lambda=1$).
Finally, the Kaiser window is defined using the zeroth‑order modified Bessel function of the first kind $I_0$ as
\begin{align}
    \label{eq:def_Kaiser}
    W_{\lambda}(s) = \mathcal{N} \left[ I_0\left(2 \lambda \sqrt{s(1-s)}\right) - I_0(0) \right].
\end{align}
Here, $\mathcal{N}$ is a normalisation factor. A visualisation of all three window functions is shown in Fig.~\ref{fig:WindowFunctions}(a).

As discussed above, the squared norm of the Fourier transform of a window function defines the signal, Eq.~\eqref{eq:Signal}.
This signal approximates the infidelity of the spectator Eq.~\eqref{eq:InfidelityXpihSignal}.
We visualise the signal for the different window functions in Fig.~\ref{fig:WindowFunctions}(b).

We now discuss the signals shown in Fig.~\ref{fig:WindowFunctions}(b) in more detail. In particular, we extract the threshold $x_{\Xpih}$. Recall that $x_{\Xpih}$ is defined such that, for all values $\nu T_{\Xpih}>x_{\Xpih}$, the infidelity remains below a target value $I_t$. To be consistent with the main manuscript, we set $I_t=10^{-5}$, as indicated by the black dashed line.

The Hann window has no free parameters. Its signal decays rapidly, resulting in a relatively small threshold value $x_{\Xpih}(10^{-5})\approx29.2$. For $\lambda=1$, the Tukey window is identical to the Hann window. For smaller values of $\lambda$ (here $\lambda=0.2$), the signal of the Tukey window decays more slowly than that of the Hann window, leading to a larger threshold value. Specifically, for $\lambda=0.2$, we obtain $x_{\Xpih}(10^{-5})\approx81.0$.

Finally, the signal of the Kaiser window decays comparatively rapidly. Here, we choose $\lambda=7.4$, which results in a small threshold value $x_{\Xpih}(10^{-5})\approx15.3$. With these parameter choices, the peak values of the normalised windows for the Hann, Tukey, and Kaiser windows are $W_{\max}\approx2.0, 1.1$, and $2.2$, respectively. These values are summarised in Tab.~\ref{tab:WindowParamsXpih}.
\begin{table}[h!]
\begin{center}
\begin{tabular}{ |c|c|c|c| } 
  \hline
    Window & Parameter     & $x_{\Xpih}$   & $W_{\max}$ \\ 
  \hline
    Hann   & --            & $\approx29.2$ & 2.0        \\ 
    Tukey  & $\lambda=0.2$ & $\approx81.0$ & $\approx1.1$     \\ 
    Kaiser & $\lambda=7.4$ & $\approx15.3$ & $\approx2.2$ \\ 
 \hline
\end{tabular}
\end{center}
\caption{Summary of window‑function parameters used to execute an $\Xpih$ gate.}\label{tab:WindowParamsXpih}
\end{table}

\subsubsection{Execution times}

\begin{figure}
    \centering
    \includegraphics[width=0.5\textwidth]{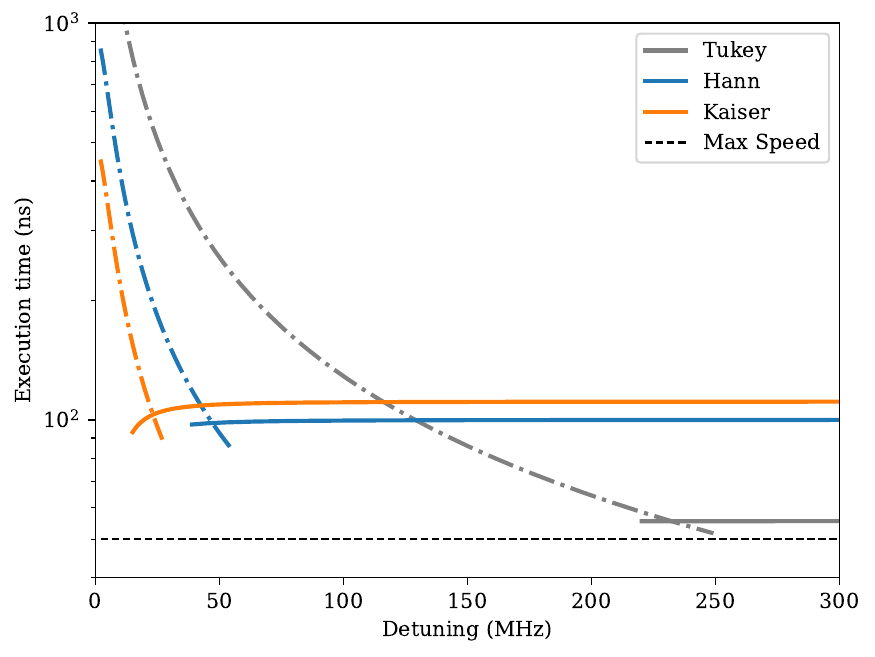}
    \caption{
    Lower bounds on the execution time of the $\Xpih$ gate as a function of detuning $\Delta/h$ for the Hann window (blue), the Kaiser window (orange), and the Tukey window (grey), using the parameters from Tab.~\ref{tab:WindowParamsXpih} and $\Omega_{\max}/h=5$ MHz. At small detunings, the requirement to minimise crosstalk while maintaining a high fidelity of the spectator qubit leads to longer execution times, as given by the lower bound $T_{\Xpih}^{\min, \mathrm{fid}}$ of Eq.~\eqref{eq:GateTimeFidelityXpih} (dash‑dotted lines). At large detunings, the execution time is set by the requirement to achieve a sufficient pulse area to ensure high fidelity of the target qubit, as given by the lower bound $T_{\Xpih}^{\min, \mathrm{area}}$ of Eq.~\eqref{eq:GateTimeAreaXpih} (solid lines). The black dashed line indicates the theoretically fastest execution time given by Eq.~\eqref{eq:XpihFastest}.
     }\label{fig:GateTimesXpih}
\end{figure}

Using the parameters listed in Tab.~\ref{tab:WindowParamsXpih} and $\Omega_{\max}/h=5$ MHz, we calculate lower bounds on the execution times of $\Xpih$ gates implemented using crosstalk‑mitigating shaped pulses. 
We particularly evaluate the lower bounds $T_{\Xpih}^{\min, \mathrm{area}}$ and $T_{\Xpih}^{\min, \mathrm{fid}}$, given by Eq.~\eqref{eq:GateTimeAreaXpih} and Eq.~\eqref{eq:GateTimeFidelityXpih}, respectively.
The results are shown in Fig.~\ref{fig:GateTimesXpih} for the Hann window (blue), the Kaiser window (orange), and the Tukey window (grey).

For large detuning values (\textit{i.e.}, in the presence of a micromagnet), the execution time of the $\Xpih$ gate is set by the pulse‑area condition, Eq.~\eqref{eq:GateTimeAreaXpih} (solid lines). As expected, the execution times obtained using shaped pulses are longer than the fastest possible execution times predicted by Eq.~\eqref{eq:XpihFastest}. The increased pulse duration is well explained by Eq.\eqref{eq:GateTimeAreaScaledXpih}. In particular, at large detuning values, such as those encountered in the presence of a micromagnet, a Tukey window may yield the fastest pulses. In particular, for detuning values $\Delta/h=100$ and $300$ MHz using a Hann or Tukey window, respectively, results in a fast execution time of $T_{\Xpih} \ge 100$ and $T_{\Xpih} \ge 56$ nanoseconds, respectively. Both values are reported in Tab.~\ref{tab:GateExecutionTimes}.

For small detuning values (\textit{i.e.}, in the absence of a micromagnet), the execution time of the $\Xpih$ gate is set by the fidelity condition, Eq.~\eqref{eq:GateTimeFidelityXpih} (dashed lines). In this regime, the gate operation must be slowed down to minimise crosstalk. In particular, for detuning values $\Delta/h=2$ and $8$ MHz using a Kaiser window, achieves an execution time $T_{\Xpih} \ge 453$ and $T_{\Xpih} \ge 259$ nanoseconds, respectively. Both values are reported in Tab.~\ref{tab:GateExecutionTimes}.

\section{Execution time of the $\CZ$-gate}

In this section, we consider the execution times of the $\CZ$ gate. To this end, we analyse the Heisenberg chain under an exchange pulse. We show that the exchange pulse implements a unitary evolution corresponding to an $\RZZ$ gate, along with unwanted $\SWAP$-like oscillations in the singlet–triplet subspace. The problem of compensating for these unwanted $\SWAP$ oscillations is formally identical to mitigating crosstalk on a spectator qubit. This observation allows us to apply the same mitigation strategies introduced in the previous section.

\subsection{Time-evolution operator of the $\CZ$-gate}

We begin by considering two spin qubits in the laboratory frame, described by the Hamiltonian
\begin{align}
H(t)=-\frac{1}{2}B_1Z_1-\frac{1}{2}B_2Z_2+\frac{1}{4}J(t)\left[X_1 X_2 + Y_1 Y_2 + Z_1 Z_2\right].
\end{align}
Here, $J(t)$ describes an exchange pulse in the time interval $t\in[t_0,t_0+T]$ of duration $T$.
These spins obey the Schr{\"o}dinger equation
\begin{align}
i\hbar\partial_t \psi(t)= H(t)\psi(t).
\end{align}
As before, we transform the Schr{\"o}dinger equation into the qubit frame via
\begin{align}
\psi'(t)=R(t)\psi(t),
\end{align}
with
\begin{align}
R(t)=\exp(-i\frac{B_1 Z_1 + B_2 Z_2}{2\hbar}[t-t_0]).
\end{align}
This yields the Schr{\"o}dinger equation for the state $\psi'(t)$ in the rotated frame,
\begin{align}
i\hbar \partial_t \psi'(t)= H'(t)\psi'(t),
\end{align}
where
\begin{align}
H'(t)=\frac{J(t)}{4}R(t)\left[X_1 X_2 + Y_1 Y_2 + Z_1 Z_2\right]R^{-1}(t).
\end{align}
The qubit is defined in this frame.

Next, we expand the Hamiltonian in the qubit basis
\begin{align}
    \ket{1} &= \ket{0_2,0_1},\\
    \ket{2} &= \ket{0_2,1_1},\\
    \ket{3} &= \ket{1_2,0_1},\\
    \ket{4} &= \ket{1_2,1_1}.
\end{align}
Here, the subscript denotes the state of the $i$th qubit. The resulting matrices are
\begin{align}
    &R(t)
    =
    \begin{bmatrix}
      \exp(-i\frac{B_1+B_2}{2\hbar}[t-t_0]) & & & \\
      & \exp(-i\frac{-B_1+B_2}{2\hbar}[t-t_0]) & & \\
      & & \exp(-i\frac{B_1-B_2}{2\hbar}[t-t_0]) & \\
      & & & \exp(-i\frac{-B_1-B_2}{2\hbar}[t-t_0]) \\
    \end{bmatrix},\\
    &X_1 X_2 + Y_1 Y_2 + Z_1 Z_2
    =
    \begin{bmatrix}
        1 &    &    &   \\
          & -1 &  2 &   \\
          &  2 & -1 &   \\
          &    &    & 1 \\
    \end{bmatrix}.
\end{align}
Introducing the detuning $\Delta := B_2-B_1$, we obtain
\begin{align}
    H'(t)
    &= \frac{J(t)}{4}
    \begin{bmatrix}
        1 &    &    &   \\
          & -1 &  2\exp(-i\frac{\Delta}{\hbar}[t-t_0]) &   \\
          &  2\exp(i\frac{\Delta}{\hbar}[t-t_0]) & -1 &   \\
          &    &    & 1 \\
    \end{bmatrix}.
\end{align}
We now solve for the time‑evolution operator $U'(t, t_0)$, which satisfies
\begin{align}
    \label{eq:SchroedingerCZ}
    i\hbar\partial_t U'(t,t_0)= H'(t)U'(t,t_0),
    \quad \text{with:} \quad
    U'(t_0, t_0)= 1.    
\end{align}
To this end, we define the phase
\begin{align}
    \label{eq:ThetaCZ}
    \theta(t) = \int_{t_0}^{t} \text{d}\tau \frac{J(\tau)}{2\hbar},
\end{align}
and introduce the ansatz
\begin{align}
    U'(t,t_0) := \RZZ(\theta(t)) \bar{K}(t,t_0).
\end{align}
Here, $\bar{K}(t,t_0)$ has the matrix representation
\begin{align}
    \bar{K}(t,t_0) 
    := 
    \begin{bmatrix}
    1 &    &    &   \\
      & K_{11}(t,t_0) &  K_{12}(t,t_0) & \\
      & K_{21}(t,t_0) &  K_{22}(t,t_0) & \\
      &               &                & 1 \\
    \end{bmatrix},
\end{align}
where $K(t,t_0)$ is a $2\times2$ matrix acting on the singlet–triplet subspace. The operator $\RZZ(\theta(t))$ commutes with both $\bar{K}(t,t_0)$ and $H'(t)$.
Substituting the ansatz into Eq.~\eqref{eq:SchroedingerCZ} immediately solves the Schrödinger equation for the states $\ket{1}$ and $\ket{4}$. Identifying $\ket{0_2,1_1} \equiv \ket{0}$ and $\ket{1_2,0_1} \equiv \ket{1}$, the Schr{\"o}dinger equation in the singlet–triplet subspace becomes
\begin{align}
    i\hbar\partial_t K(t,t_0)
    = \frac{J(t)}{2} R_z(\Delta [t-t_0]/\hbar) X R_z(-\Delta [t-t_0]/\hbar) K(t,t_0)
    \quad \text{with:} \quad
    K(t_0, t_0)= I.
\end{align}
Identifying $\Omega(t)\equiv J(t)$ and noting that $\varphi=0$, this equation is formally identical to the Schr{\"o}dinger equation governing the spectator qubit in the crosstalk problem (\textit{cf.}~Eq.~\eqref{eq:SchroedingerSpectatorQubit}). We may therefore apply the same solution method. To this end, we redefine the relevant quantities as
\begin{align}
    \label{eq:defs_gamma_CZ}
    \gamma_t &= \frac{\Delta}{\hbar}[t-t_0], \\
    \label{eq:defs_alpha_CZ}
    \alpha_t &= -\sigma_{\Delta} \arctan(\frac{J(t)}{|\Delta|}), \\
    \label{eq:defs_Delta_CZ}
    \Delta_t' &= -\sigma_{\Delta} \sqrt{\Delta^2+J^2(t)}, \\
    \label{eq:defs_beta_CZ}
    \beta_t' &= \int_{t_0}^{t}\text{d}\tilde{t}\frac{\Delta_{\tilde{t}}'}{\hbar}.
\end{align}
By direct analogy, the formal solution is then given by (\textit{cf.}~Eq.~\eqref{eq:SpectatorUnitary})
\begin{align}
    \label{eq:SingletTripletUnitary}
    K(t,t_0) = 
    R_z(\gamma_t) R_y(\alpha_t) R_z(\beta_t)
    U_4(t,t_0) R_z(-\beta_{t_0}) R_y(-\alpha_{t_0}) R_z(-\gamma_{t_0}).
\end{align}

\subsection{Fidelity of the $\CZ$-gate}

Our goal is to identify pulses such that $U'(t,t_{0})$ implements a $\CZ$ gate with high fidelity. This objective is achieved if $U'(t,t_{0})$ reproduces a $\CZ$ gate up to a global phase $\chi$ and the application of virtual $\VZ$-rotations on each qubit. Accordingly, our target unitary is
\begin{align}
    U_i 
    := 
    \CZ \cdot R_{z_1}(\varphi_1) \otimes R_{z_2}(\varphi_2) \cdot \exp(i\frac{\chi}{2}).
\end{align}
To reproduce this unitary from $U'(t_0+T_{\CZ},t_0) = \RZZ(\theta(t_0+T_{\CZ})) \bar{K}(t_0+T_{\CZ},t_0)$ with high fidelity, the gate time $T_{\CZ}$ is chosen such that
\begin{align}
    \label{eq:PulseAreaCZ}
    \frac{\pi}{2} 
    =
    \theta(T_{\CZ}).
\end{align}
Note that the value of $\theta(T_{\CZ}=\pi/2)$ can be derived from the equations which follow below. We fix this phase here upfront, to simplify the exposition.
As before, the pulse is engineered such that $K(t_0+T_{\CZ},t_0)$ is close to the identity, up to conjugation by known $Z$-rotations and a phase $\zeta$,
\begin{align}
    K(t_0+T_{\CZ},t_0) = R_z(\eta)\exp(-i\frac{P_\epsilon}{2}|\epsilon|)\exp(i\zeta)R_z(-\xi).
\end{align}
For unitaries of this form, one finds
\begin{align}
    &\mathrm{Tr}\left[U_i^\dagger U'(t_0+T_{\CZ},t_0)\right] 
    \nonumber\\
    &=
    \mathrm{Tr}\left[
    \begin{pmatrix}
    e^{i\frac{\varphi_1+\varphi_2-\chi-\frac{\pi}{2}}{2}} &    &    &   \\
      & e^{i\frac{\varphi_2-\varphi_1-\chi+\frac{\pi}{2}-\eta+\xi+2\zeta}{2}} &   & \\
      &  & e^{i\frac{\varphi_1-\varphi_2-\chi+\frac{\pi}{2}+\eta-\xi+2\zeta}{2}} & \\
      &  &  & -e^{-i\frac{\varphi_1+\varphi_2+\chi+\frac{\pi}{2}}{2}} \\
    \end{pmatrix}
    \begin{pmatrix}
     1 &   &   \\
       &  \exp(-i\frac{P_{\epsilon}}{2}|\epsilon|) &   \\
       &   & 1 \\
    \end{pmatrix}
    \right].
\end{align}
By choosing the global phase and the virtual $Z$-rotations such that
\begin{align}
    \chi &= \frac{\pi}{2} \\
    \varphi_2 + \varphi_1 &=\pi, \\
    \zeta_n &= n \pi, \\
    [\varphi_2 - \varphi_1]_n &=\eta-\xi+2n\pi,
\end{align}
we find that $U(t_0+T_{\CZ},t_0)$ implements a $\CZ$-gate with fidelity
\begin{align}
    F_{\CZ}
    =
    \frac{|\mathrm{Tr}[U_i^{\dagger} U'(T_{\CZ}+t_0,t_0)]|^2+4}{4(4+1)}
    = 
    \frac{|2+\mathrm{Tr}[I\cos(|\epsilon|/2) 
    -i P_\epsilon\sin(|\epsilon|/2)]|^2+4}{4(4+1)}.
\end{align}
Evaluating the trace yields
\begin{align}
    F_{\CZ}
    = 
    \frac{|1+\cos(|\epsilon|/2)|^2+1}{(4+1)}
    = 
    \frac{2 + 2\cos(|\epsilon|/2) + \cos^2(|\epsilon|/2)}{(4+1)}.
\end{align}
Conversely, the infidelity of the $\CZ$ gate is given by
\begin{align}
    \label{eq:InfidelityCZ}
    I_{\CZ}:=1-F_{\CZ}
    = 
    \frac{2 - 2\cos(|\epsilon|/2) + \sin^2(|\epsilon|/2)}{5}
    \approx
    \frac{|\epsilon|^2}{10}.
\end{align}

\subsection{Execution time of the $\CZ$ gate}

The results in this section are a straightforward adaptation of the single-qubit case.

\subsubsection{Fastest possible gate time}

As before, using the pulse‑area condition, Eq.~\eqref{eq:PulseAreaCZ}, and the fact that $J(t)\le J_{\max}$, we obtain a lower bound on the gate‑execution time,
\begin{align}
    \label{eq:ExecutionTimeCZFastest}
    T_{\CZ} 
    \ge 
    T_{\CZ}^{\min}
    :=
    \frac{h}{2 J_{\max}}.
\end{align}
In our case, we assume $J_{\max}/h=10$MHz. Consequently, the fastest possible execution time of a $\CZ$ gate is $50$ nanoseconds.

\subsubsection{Synchronisation}

\begin{figure}
    \centering
    \includegraphics[width=0.5\textwidth]{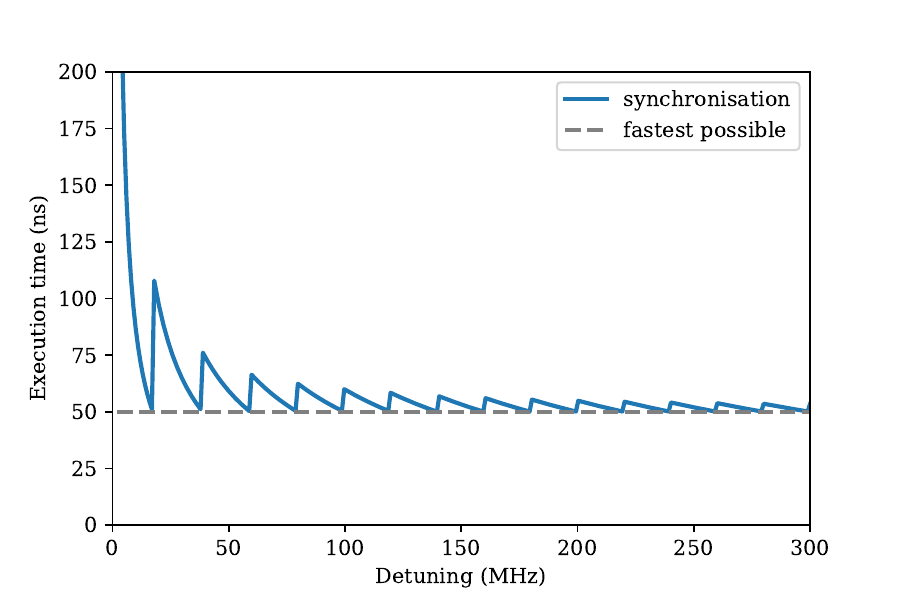}
    \caption{
    Execution time of the $\CZ$ gate [Eq.~\eqref{eq:ExecutionTimeCZSynch}, blue line] as a function of detuning $\Delta$, assuming $J_{\max}/h = 10$ MHz and the use of the synchronisation method. The dashed line indicates the fastest possible execution time given by Eq.~\eqref{eq:ExecutionTimeCZFastest}.
    }\label{fig:CZSynch}
\end{figure}

As in the single‑qubit case, applying a rectangular pulse with constant exchange coupling $J$ yields the time‑evolution operator, Eq.~\eqref{eq:SingletTripletUnitary},
\begin{align}
    K(t_0+T_{\CZ},t_0) = 
    R_z(\gamma_{t_0+T_{\CZ}}) R_y(\alpha_{t_0+T_{\CZ}}) 
    R_z(\beta_{t_0+T_{\CZ}}-\beta_{t_0}) 
    R_y(-\alpha_{t_0}) 
    R_z(-\gamma_{t_0}).
\end{align}
The outer $R_z$ rotations can be absorbed into virtual gates. Since $\alpha_{t_0+T_{\CZ}} = \alpha_{t_0} \neq 0$, we require the central rotation $R_z(\beta_{t_0 + T_{\CZ}} - \beta_{t_0})$ to be equal to the identity (or minus the identity). This condition is satisfied if
\begin{align}
 |\beta_{t_0+T_{\CZ}}-\beta_{t_0}|=2\pi m,
\end{align}
which leads to the gate‑execution time, when using  Eq.~\eqref{eq:defs_Delta_CZ} and Eq.~\eqref{eq:defs_beta_CZ},
\begin{align}
    T_{\CZ}(m) = \frac{hm}{\sqrt{\Delta^2+J^2}}.
\end{align}
At the same time, Eq.~\eqref{eq:PulseAreaCZ} must yield the correct pulse area, such that $T_{\CZ}$ must also satisfy
\begin{align}
    T_{\CZ} = \frac{h}{2J}.
\end{align}
For both conditions to be synchronised, the exchange amplitude must be chosen as
\begin{align}
    J(m) = \frac{|\Delta|}{\sqrt{4m^2-1}}.
\end{align}
This results in feasible gate-execution times
\begin{align}
T_{\CZ}(m) = \frac{h}{2|\Delta|}\sqrt{4m^2-1}.
\end{align}
Imposing the constraint $J(m)\le J_{\max}$ leads to
\begin{align}
    m\ge\frac{\sqrt{\Delta^2+J_{\max}^2}}{2 J_{\max}}.
\end{align}
Accordingly, the minimal viable integer $m$ is
\begin{align}
    m_{\min}
    =
    \bigg \lceil \frac{ \sqrt{\Delta^2 + J_{\max}^2}}{2 J_{\max}}\bigg\rceil.
\end{align}
The fastest possible $\CZ$-gate execution time achievable via synchronisation is therefore
\begin{align}
    \label{eq:ExecutionTimeCZSynch}
  T_{\CZ}^{\min,\mathrm{syn}} 
  = 
  \frac{h}{2|\Delta|}\sqrt{4    \bigg \lceil \frac{ \sqrt{\Delta^2 + J_{\max}^2}}{2 J_{\max}}\bigg\rceil^2-1}
  \ge
  \frac{h}{2J_{\max}}=T_{\CZ}^{\min}.
\end{align}
A visualisation of $T_{\CZ}^{\min,\mathrm{syn}}$ for $J_{\max}/h=10$ MHz and $\Delta/h\in[2,300]$ MHz is shown in Fig.~\ref{fig:CZSynch}. The execution time ranges from $50$ to $433$ nanoseconds and becomes particularly long for small detuning values $\Delta$, as encountered in devices without a micromagnet. For a detuning value of $\Delta/h=8$ MHz, which is widely used in the main manuscript, we predict a gate-execution time of approximately $108$ nanoseconds.

\subsubsection{Shaped pulses}

To estimate the execution times of the $\CZ$ gate using shaped pulses, we employ the same framework as before, replacing $\Omega(t)$ by $J(t)$. While residual exchange can sometimes be significant, we assume that $J(t_0), J(t_0+T_{\CZ})\ll|\Delta|$, such that $\alpha_{t_0}=\alpha_{t_0+T_{\CZ}}\approx0$ according to Eq.\eqref{eq:defs_alpha_CZ}.
Under this assumption, the time-evolution operator in Eq.~\eqref{eq:SingletTripletUnitary} becomes
\begin{align}
    K(T_{\CZ}+t_0,t_0) = 
    R_z(\gamma_{T_{\CZ}+t_0} + \beta_{T_{\CZ}+t_0})
    U_4(T_{\CZ}+t_0,t_0)
    R_z(-\beta_{t_0}-\gamma_{t_0}).
\end{align}
Using the same method as before, we introduce a dilated time variable
\begin{align}
    \label{eq:def_dilated_time_CZ}
    s(t) := \frac{-\sigma_{\Delta}\beta_t}{\nu T_{\CZ}}
    \quad \text{such that:}
    \quad s(t_0)=0
    \quad \text{and}
    \quad s(t_0+T_{\CZ})=1.
\end{align}
Applying a first‑order Magnus expansion for the resulting Schr{\"o}dinger equation, one again finds
\begin{align}
    U_4(T_{\CZ}+t_0,t_0) = \tilde{U}_4(1)=\exp(-\frac{|\epsilon|}{2}P_{\epsilon}),
\end{align}
with
\begin{align}
    \epsilon 
    := 
    i\int_{0}^1\mathrm{d}s     \dot{\tilde{\alpha}}_s \exp(-i\sigma_{\Delta}\nu T_{\CZ} s).
\end{align}
Using a window‑function pulse,
\begin{align}
    \label{eq:def_dilated_pulse_CZ}
    \tilde{J}(s) := J(t(s)) = |\Delta|\tan\left[AW(s)+\arctan(\frac{J(t_0)}{|\Delta|})\right]\ge0,
\end{align}
and following the same reasoning as in the single-qubit case, one obtains
\begin{align}
    \label{eq:EpsilonAndSignalCZ}
    |\epsilon|^2 = |\nu T_{\CZ} A|^2 S[W(\nu T_{\CZ})],
\end{align}
where the signal is defined as
\begin{align}
    \label{eq:SignalCZ}
    S[W(\nu T_{\CZ})] := \left| \int_{0}^1\mathrm{d}s  
    W(s) \exp(i\nu T_{\CZ} s)\right|^2.
\end{align}
The modified time‑dilation and pulse‑area conditions become [cf.~Eqs.~\eqref{eq:TimeDilationCondition} and~\eqref{eq:PulseAreaCondition}]
\begin{align}
    \label{eq:TimeDilationConditionCZ}
    1 = \frac{\nu \hbar}{|\Delta|} \int_{0}^{1} \mathrm{d}s \cos(A W(s) + \arctan(\frac{J(t_0)}{|\Delta|}) ),\\
    \label{eq:PulseAreaConditionCZ}
    \pi 
    = 
    \int_{t_0}^{t_0+T_{\CZ}}\frac{J(t)}{\hbar}\mathrm{d}t 
    = 
    \nu T_{\CZ} \int_{0}^{1} \mathrm{d}s
    \sin(A W(s)+\arctan(\frac{J(t_0)}{|\Delta|})).
\end{align}
Using the same arguments as before, the pulse-area condition yields a lower bound on the execution time,
\begin{align}
    \label{eq:GateTimeAreaCZ}
    T_{\CZ}
    \ge 
    T_{\CZ}^{\min, \mathrm{area}}
    :=
    \frac{h}{2J_{\max}} W_{\max}
    \mathrm{sinc}\left(\arctan(\frac{J_{\max}}{|\Delta|})\right)
    \stackrel{J_{\max}\ll|\Delta|}{\approx}
    T_{\CZ}^{\min}W_{\max}.
\end{align}
Furthermore, by introducing a threshold $x_{\CZ}(I_t)$ such that
\begin{align}
    \forall \nu T_{\CZ} \ge x_{\CZ}(I_t): 
    \quad
    S[W(\nu T_{\CZ})]\le \frac{40}{\pi^4}I_t,
\end{align}
we ensure that, for all $\nu T_{\CZ} \ge x_{\CZ}(I_t)$,
\begin{align}
    I_{CZ}
    \approx
    |\nu T_{\CZ} A|^2
    \frac{S[W(\nu T_{\CZ})]}{10}
    = 
    \left|
    \frac{A \pi}{\int_{0}^{1}\sin(A W(s))}
    \right|^2
    \frac{S[W(\nu T_{\Xpih})]}{10}
    \le
    \left|\frac{\pi^2}{2}\right|^2
    \frac{S[W(\nu T_{\Xpih})]}{10}
    \le
    I_t.
\end{align}
This threshold again yields a lower bound on the execution time of the $\CZ$ gate
\begin{align}
    \label{eq:GateTimeFidelityCZ}
    T_{\CZ}
    \ge
    T_{\CZ}^{\min, \mathrm{fid}}
    :=
    \frac{x_{\CZ}(I_t) h}{2\pi \sqrt{\Delta^2+J_{\max}^2}}.
\end{align}

\begin{figure}
    \centering
    \includegraphics[width=0.9\textwidth]{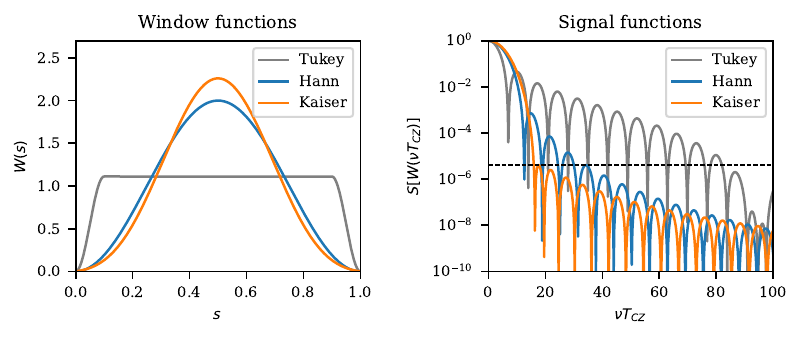}
    \caption{
    (a) Hann (blue), Kaiser (orange), and Tukey (grey) window functions using the parameters listed in Tab.~\ref{tab:WindowParamsCZ}.\ (b) Corresponding signal functions $S[W(\nu T_{\CZ})]$. The dashed line indicates the threshold value $40 I_t/\pi^4$ for a target infidelity of $I_t=10^{-5}$.}\label{fig:WindowFunctionsCZ}
\end{figure}

\begin{figure}
    \centering
    \includegraphics[width=0.5\textwidth]{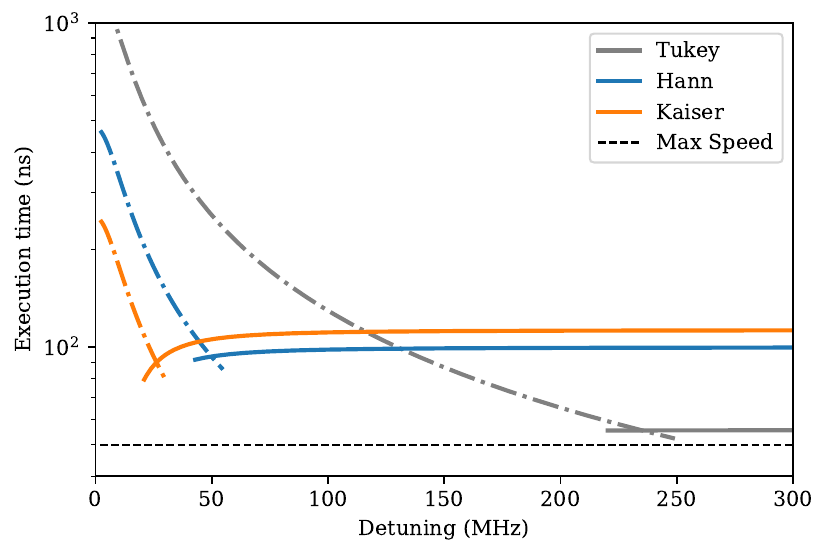}
    \caption{
    Lower bounds on the execution time of the $\CZ$ gate as a function of detuning $\Delta/h$ for the Hann (blue), the Kaiser (orange), and the Tukey window (grey), using the parameters listed in Tab.~\ref{tab:WindowParamsCZ} and $J_{\max}/h=10$ MHz. At small detunings, the need to minimise unwanted $\SWAP$-like oscillations in the singlet-triplet subspace increases the execution time, as given by Eq.~\eqref{eq:GateTimeFidelityCZ} (dash‑dotted lines). At large detunings, the execution time is dominated by the requirement to achieve a sufficient pulse area, as described by Eq.~\eqref{eq:GateTimeAreaCZ} (solid lines). The black dashed line indicates the theoretically fastest execution time given by Eq.~\eqref{eq:ExecutionTimeCZFastest}.
    }\label{fig:GateTimesCZ}
\end{figure}

\textit{Examples:---}To illustrate these bounds, we again consider the Hann, Tukey, and Kaiser windows defined in Eqs.~\eqref{eq:def_Hann}--\eqref{eq:def_Kaiser}. The parameters used are listed in Tab.~\ref{tab:WindowParamsCZ}. Visualisations of the window functions and their corresponding signals are shown in Fig.~\ref{fig:WindowFunctionsCZ}, from which the values of $W_{\max}$ and $x_{\CZ}$ can be extracted.
\begin{table}[]
\begin{center}
\begin{tabular}{ |c|c|c|c| } 
  \hline
    Window & Parameter & $x_{\CZ}$   & $W_{\max}$ \\ 
  \hline
    Hann   & --  & $\approx29.9$ & 2.0 \\ 
    Tukey  & $\lambda=0.2$ & $\approx81.9$ & $\approx1.1$     \\ 
    Kaiser & $\lambda=7.67$ & $\approx15.83$ & $\approx2.26$ \\ 
 \hline
\end{tabular}
\end{center}
\caption{Summary of window-function parameters used to execute a CZ gate.
}\label{tab:WindowParamsCZ}
\end{table}
Using the parameters in Tab.~\ref{tab:WindowParamsCZ}, the execution-time bounds given by Eqs.~\eqref{eq:GateTimeAreaCZ} and~\eqref{eq:GateTimeFidelityCZ} are evaluated and shown in Fig.~\ref{fig:GateTimesCZ} as a function of the detuning $\Delta/h$.

At large detunings $|\Delta|\gg J_{\max}$ (\textit{i.e.}, if a micromagnet is present), the pulse‑area condition [Eq.~\eqref{eq:GateTimeAreaCZ}, solid lines] dominates. Specifically, for detuning values $\Delta/h=100$ and $300$ MHz using a Hann or Tukey window, respectively, achieves a fast execution time of $T_{\CZ} \ge 98$ and $T_{\CZ} \ge 55$ nanoseconds. Both values are reported in Tab.~\ref{tab:GateExecutionTimes}.

At small detuning values $|\Delta| \lesssim J_{\max}$ (\textit{i.e.}, if a micromagnet is absent), the fidelity constraint [Eq.~\eqref{eq:GateTimeFidelityCZ}, dash‑dotted lines] sets the lower bound. In this regime, the gate operation must be slowed down to minimise undesired $\SWAP$ oscillations. In particular, for detuning values $\Delta/h=2$ and $8$ MHz, respectively, using a Kaiser pulse, achieves an execution time of $T_{\CZ} \ge 247$ and $T_{\CZ} \ge 196$ nanoseconds. Both values are reported in Tab.~\ref{tab:GateExecutionTimes}.

\section{Execution time of the resonant SWAP gate}

Following Ref.~\cite{Rimbach-Russ2023Sep}, the pulse-area condition for the resonant SWAP gate is
\begin{align}
    \label{eq:PulseAreaSWAP}
    \pi 
    =
    \theta(T_{\SWAP})
    =:
    \int_{t_0}^{t_0+T_{\SWAP}} \text{d}\tau \frac{J(\tau)}{2\hbar}.
\end{align}
Using the bound $J(t)\le J_{\max}$, we obtain a lower bound on the gate‑execution time,
\begin{align}
    \label{eq:ExecutionTimeSWAPFastest}
    T_{\SWAP}\ge T_{\SWAP}^{\min}:=\frac{h}{ J_{\max}}.
\end{align}
In our case, we assume $J_{\max}/h=10$~MHz. Consequently, the fastest possible execution time of a resonant SWAP gate is $100$ nanoseconds.
\FloatBarrier
\bibliography{references}